\documentclass[aps,pre,twocolumn,groupedaddress,showpacs,floatfix,amsmath]{revtex4}
\usepackage{graphicx,amsmath}
\usepackage{color}
\begin{document}
\preprint{}
\title{Activity-induced clustering in model dumbbell swimmers: \\ The role of hydrodynamic interactions}
\author{Akira Furukawa$^{1}$, Davide Marenduzzo$^2$, and Michael E Cates$^2$} 
\affiliation{$^1$Institute of Industrial Science, University of Tokyo, Meguro-ku, Tokyo 153-8505, Japan \\$^2$ School of Physics and Astronomy, University of Edinburgh, JCMB Kings Buildings, Mayfield Road, Edinburgh EH9 3JZ, United Kingdom} 
\date{Received: date / Revised version: date}
\begin{abstract} 
Using a fluid-particle dynamics approach, we numerically study the effects of hydrodynamic interactions on the collective dynamics of active suspensions within a simple model for bacterial motility: each microorganism is modeled as a stroke-averaged dumbbell swimmer with prescribed dipolar force pairs. 
Using both simulations and qualitative arguments, we show that, when the separation between swimmers is comparable to their size, the swimmers' motions are strongly affected by activity-induced hydrodynamic forces. 
To further understand these effects, we investigate semi-dilute suspensions of swimmers in the presence of thermal fluctuations.  A direct comparison between simulations with and without hydrodynamic interactions shows these to enhance the dynamic clustering at a relatively small volume fraction; with our chosen model the key ingredient for this clustering behavior is hydrodynamic trapping of one swimmer by another, induced by the active forces. 
Furthermore, the density dependence of the motility (of both the translational and rotational motions) exhibits distinctly different behaviors with and without hydrodynamic interactions; we argue that this is linked to the clustering tendency. 
Our study illustrates the fact that hydrodynamic interactions not only affect kinetic pathways in active suspensions, but also cause major changes in their steady state properties. 
\end{abstract}
\pacs{47.63.Gd, 87.18.Gh, 82.70.-y, 47.57.J-}
\maketitle

\section{Introduction}
It is well established that various microorganisms, such as bacteria and algae, propel themselves through a suspending medium using a nonreciprocal cyclic motion \cite{ChildlessB,LighthillB}. Despite the progress that has been made in elucidating the self-propulsion mechanisms of various microorganisms in isolation, an understanding of their collective dynamics is still elusive. Recent experimental and simulation studies have shown that complicated interactions arising from the microorganisms' activity produce diverse nonequilibrium cooperative phenomena, whose behaviors are notably different from those observed in passive systems. Such fascinating aspects of active systems are suggestive of new underlying principles; the search for these is the current focus of intense study among the soft matter physics community (see recent reviews \cite{RamaswamyR,CatesR,MarchettiR} and references therein).  

Hydrodynamic interactions (HIs), and their dynamic coupling to activity, are thought to be among the key elements that govern transport and rheological properties in suspensions of swimming microorganisms (see \cite{Lauga_PowersR} for a review). The hydrodynamic interactions between a pair of microorganisms have been extensively studied and their general features have been well revealed in certain simple situations; for example, when two microorganisms are far apart compared to their sizes, a weak attractive interaction acts between them. This is because the dominant contribution is of a dipole-dipole character (which is attractive for particles that are free to rotate). 

However, we are still far from a thorough understanding of the role of HIs in nondilute systems of active particles. When the distances between microorganisms are comparable to their sizes, HIs in the near field become important, but the effects of such interactions are not easy to understand in general. This is because the details of both the swimming mechanisms and the particle shapes come into play with increased proximity (where not just dipolar terms but higher order multipoles contribute), so that the HIs among microorganisms at moderate or high density are more complicated and less universal than the dilute case. Thus, it is interesting and important to examine many-body HIs through simulations of minimal microorganism models. From such models it may be possible to extract some generic features of cooperative phenomena in active suspensions of various types. 

It is well recognized that, in a wide variety of systems, activity produces marked dynamic interactions, resulting in collective fluctuations or structure formation. Examples include giant number fluctuations \cite{Simha_Ramaswamy,Narayan_Ramaswamy_Menon,Deseigne_Dauchot_Chate}, clustering (or swarming) \cite{Schaller_Weber_Semmrich_Frey_Bausch,Theurkauff_Cottin-Bizonne_Palacci_Ybert_Bocquet}, and bulk phase separation \cite{Tailleur_Cates,Cates_Tailleur,Farrell-Marchetti-Marebduzzo-Tailleur,Fily-Marchetti,Redner-Hagen-Brskaran,Stenhammar_Tiribocchi_Allen_Marrenduzo_Cates,Buttioni_Bialke_Kummel_Lowen_Bechinger_Speck,Cates-Marenduzzo-Paginabarrafa-Tailleur}. These phenomena can occur without any direct potential interactions, and therefore have a purely dynamical origin. 
In systems manifesting such a dynamic cooperativity, a fundamental question is how HIs influence or change the activity-induced interactions. 

Although certain collective effects characteristic of active systems are believed to be greatly influenced by HIs, there is still no general consensus on their roles in such phenomena. Indeed, for the reasons given above, this can depend on particle shape and/or swimming mechanism so that such questions must be asked within the context of well defined models. One popular model is that of squirmers: spherical particles with a prescribed surface velocity field \cite{Lighthill,Blake}. For squirmers in two dimensions it was found that the phase separation, otherwise caused by a collisional reduction in mean propulsion speed of swimmers at high density \cite{Stenhammar_Tiribocchi_Allen_Marrenduzo_Cates}, is switched off by HIs \cite{Fielding}. Another popular model is based on dumbbell shaped swimmers subject to a set of discrete forces acting on the two solid particles and on the fluid \cite{Graham1,Graham2,Graham3}. As we explain further below, this model may need to be supplemented by an {\it ad hoc} rule to govern the case where a second swimmer is about to occupy the spatial position at which the active force from a first swimmer acts on the fluid. Even with such a rule, the model seems a better starting point than squirmers when modeling suspensions of rod-like motile bacteria. Specifically, squirmers can only exert active torques on each other, whereas rod-like swimmers can exert both active and passive torques. Note that torques are particularly important in active systems (e.g. \cite{Fielding}) since they rotate not just the particle, but also its swimming direction.

In this study, to further understand the role of HIs in collective dynamics for nonspherical swimmers, we numerically investigate the dynamics of non-dilute active suspensions of self-propelled dumbbells in three dimensions. We especially focus on activity-induced clustering at modest concentrations \cite{Schaller_Weber_Semmrich_Frey_Bausch,Theurkauff_Cottin-Bizonne_Palacci_Ybert_Bocquet}. This can be seen as a precursor to activity-induced phase separation. However we do not address full phase separation, which can only be simulated accurately with enormous system sizes \cite{Stenhammar_Marenduzzo_Allen_Cates}, such that a study of the role of near-field HIs in active phase separation remains out of reach with current computers. (Some far-field effects might instead be represented within a continuum model \cite{Wittkowski_Tiribocchi_Stenhammar_Allen_Marenduzzo_Cates}.)
  
Our simulation method fully takes into account the solvent dynamics and thus is suitable for the investigation of HIs, as will be explained in the next section. In Sec.~III, we demonstrate that the activity-induced HIs can strongly alter both the translational and rotational dynamics at modest densities. At a relatively small volume fraction, an activity-induced hydrodynamic attraction dominates the collective dynamics of the swimmers, resulting in a stronger slowing down, and hence enhanced clustering, compared to the case without HIs. 
The volume fraction dependence of the motility is found to be much stronger with HIs than without them; this also influences the clustering tendency, as we discuss in Sec.~IV. Section V gives our conclusions and a further discussion.

\section{Numerical methods}

Many-body hydrodynamic interactions in semidilute regimes are difficult to deal with even using simulations because of their intrinsically long-range and time-dependent character; in principle we must solve the Navier-Stokes equation with moving boundary conditions at the solvent-particle interfaces. To confront these difficulties, in the past decade, several hybrid simulation techniques \cite{LBM1,LBM2,FPD1,FPD2,SPM} as well as other mesoscopic methods \cite{MPCD,DPD} for the dynamics of complex colloidal suspensions have been developed. Of these methods, we here adopt the fluid-particle-dynamics (FPD) method \cite{FPD1,FPD2} to incorporate HIs into the study of a model active suspension.

Within the framework of the FPD method, HIs can be approximately taken into account (i) by treating a rigid colloidal particle as a nondeformable but highly viscous fluid particle, and (ii) by replacing the particle-fluid boundary with a smoothed boundary. This simple scheme considerably reduces the numerical cost and the mathematical complexity of the computation even while preserving the basic nature of HIs in many-body colloidal systems. The validity of this method has been examined by several authors \cite{FPD1,FPD2,FPD3,FPD4,Fujitani,Jibuti_Rafai_Peyla}. It is good at capturing both far-field and intermediate-field aspects of the HIs, though like many other methods it cannot (at reasonable computational cost) also resolve the divergence of lubrication forces that occurs when hard suspended particles come into very close contact \cite{FPD1,FPD2}. In practice the model can be considered to describe colloids whose hard-core radius (set by the interparticle pair potential) slightly exceeds their hydrodynamic radius, so that thin lubrication films never dominate.

\subsection{Model swimmer system}

\begin{figure}[hbt] 
\includegraphics[width=0.485\textwidth]{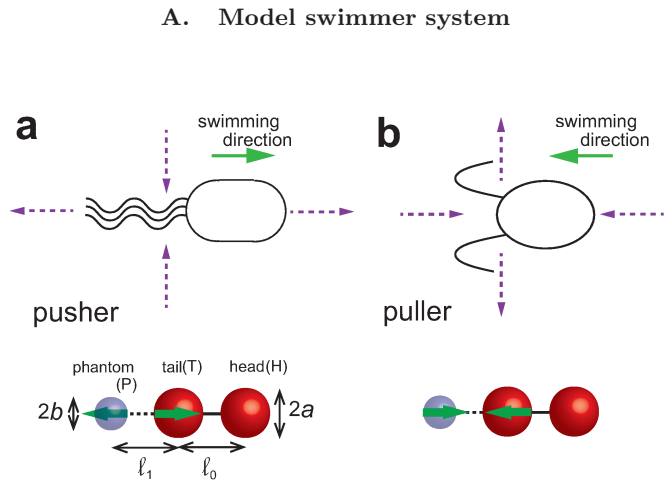}
\caption{(Color online)
Schematic of a model swimmer. Here, a swimming bacterium is represented 
by two spherical particles with radius $a_{\rm H}=a_{\rm T}=a$ and a ``phantom" spherical particle with radius $a_{\rm P}=b$. The center-to-center distance between the head and tail particles and that between the phantom and tail particles are fixed at $\ell_0$ and $\ell_1$, respectively. 
The back-to-back (a) and face-to-face (b) force configurations correspond respectively to the pusher (extensile) and puller (contractile) swimming mechanisms. 
The (green) solid and (purple) dashed arrows respectively indicate schematically the swimming direction and the flow field around the swimmer. }
\label{Fig1}
\end{figure}

Our model of a swimming microorganism comprises a dumbbell with a prescribed dipolar force pair, which is essentially the same as the models used in Refs.~\cite{Graham1,Graham2} and later in Ref.~\cite{Gyrya_Aranson_Berlyand_Karpeev}. Each dumbbell is composed of two real particles plus one ``phantom" particle -- so called because (a) it merely follows the motion of the pair of real particles to which it is attached and (b) it can overlap with the other real particles in nearby swimmers. The phantom particle can be thought of as modeling the effect of a thin flagellar bundle, whereby a force is exerted on the fluid at a position displaced from the rod-like bacterial body (represented by the dimer of real particles). 

The positions ${\mbox{\boldmath$R$}}_{m}$ and the radii $a_m$ of the particles and the distances of separation between them determine the shape of the swimmer, where the subscripts $m={\rm H}$ and ${\rm T}$ denote the head and tail particles of the dumbbell, respectively, and $m={\rm P}$ denotes the phantom particle (see Fig.~\ref{Fig1}). 
The swimmer's orientation is characterized by a unit vector $\hat{\mbox{\boldmath$n$}}=({\mbox{\boldmath$R$}}_{\rm H}-{\mbox{\boldmath$R$}}_{\rm T})/|{\mbox{\boldmath$R$}}_{\rm H}-{\mbox{\boldmath$R$}}_{\rm T}|$. The propulsion force $-f_{\rm act}\hat{\mbox{\boldmath$n$}}$ is exerted on the fluid via the phantom particle; an equal and opposite force $f_{\rm act}\hat{\mbox{\boldmath$n$}}$ is exerted on the tail particle, which ensures the dipolar character of the swimming mechanism. Hence there is no net external force on a box of fluid that fully encloses the swimmer.
Face-to-face ($f_{\rm act}>0$) and back-to-back ($f_{\rm act}<0$) force configurations correspond to ``pusher" (extensile) and ``puller" (contractile) microorganisms respectively (Fig.~1). Although, in this paper, we limit our study to the dynamics of pushers such as motile bacteria, it would be straightforward to perform a corresponding simulation for pullers of the same geometry. 

A similar dumbbell model, and its hydrodynamic instabilities, were theoretically investigated in Ref.~\cite{Baskaran_Marchetti} by analyzing hydrodynamic equations derived by a coarse-graining approach. It is one of several models, used to address active suspensions, in which surface stresses or force configurations are prescribed
\cite{Saintillan_Shelley1,Saintillan_Shelley2,Saintillan_Shelley3,Haines_Sokolov_Aranson_Berlyand_Karpeev,Haines_Aranson_Berlyand_Karpeev}. 
An alternative approach is to specify a surface slip velocity, which is usually done for spherical particles resulting in the squirmer model mentioned previously \cite{Lighthill,Blake,Ishikawa_Simmonds_Pedley}. Recently, several groups have investigated many-body HIs in squirmer suspensions \cite{Ishikawa_Pedley,Llopis_Pagonabarraga,Alarcon_Pagonabarraga,Gotze_Gompper,Fielding,Morlina_Nakayama_Yamamoto,Zottl_Stark}. Although both force-prescribed (rodlike) and velocity-prescribed (spherical) particles exhibit some similar effects, such as an enhancement of the particle diffusion, their differences in shapes and swimming mechanisms may cause important changes in the transport properties \cite{Saintillan_Shelley2}. For example, these models show distinct flow responses, largely owing to the difference in the shapes, resulting in different rheological properties \cite{Saintillan_Shelley2,Ishikawa_Pedley}. 
The intrinsic difference in the models can cause strong variations in the nature of near-field HIs as well as in the interparticle torques at close approach, both of which will alter cooperative behaviors at finite density. We will draw attention to these differences occasionally in what follows, in relation to recent simulations of squirmers \cite{Fielding}.

\subsection{Basic equations} 
 
We now briefly explain how to adapt the FPD method to the case of self-propelled dumbbells. 
The two real particles in each dimer are represented by a smooth position-dependent viscosity so that 
\begin{eqnarray} 
\eta({\mbox{\boldmath$r$}})=\eta_0+\sum_{\alpha}\sum_{m={\rm H,T}}(\eta_{\rm p}-\eta_0)\phi_m^{\alpha}({\mbox{\boldmath$r$}}), \label{viscosity}
\end{eqnarray}
where $\eta_0$ and $\eta_{\rm p}$ are the viscosities of the solvent liquid and of the (nearly) rigid particles, respectively. The phantom particle ($m={\rm P}$) does not alter the local viscosity. The function $\phi_m^{\alpha}({\mbox{\boldmath$r$}})$ represents  the smoothed profile of the $m$-th particle of the $\alpha$-th swimmer, 
\begin{eqnarray}
\phi_{m}^{\alpha}({\mbox{\boldmath$r$}})=\frac{1}{2}\biggl[1+\tanh \biggl(\frac{a_m-|{\mbox{\boldmath$r$}}-{\mbox{\boldmath$R$}}_{m}^{\alpha}|}{\xi_m}\biggr)\biggr], \label{profile}
\end{eqnarray} 
where $\xi_m$ is the interface thickness. The rigidity of the real particles is approximately sustained by the large viscosity difference, $\eta_{\rm p}/\eta_{\rm 0}\gg 1$. 

With $\eta({\mbox{\boldmath$r$}})$ obeying Eq.(\ref{viscosity}), FPD  describes the dynamics by simply solving the usual Navier-Stokes equation for the velocity field: 
\begin{eqnarray}
\rho\bigl(\frac{\partial}{\partial t}
+\mbox{\boldmath$v$}\cdot\nabla\bigr)\mbox{\boldmath$v$}
= \nabla\cdot\bigl[{\stackrel{\leftrightarrow}{\mbox{\boldmath$\sigma$}}}-p{\stackrel{\leftrightarrow}{\mbox{\boldmath$\delta$}}} +{\stackrel{\leftrightarrow}{\mbox{\boldmath$\sigma$}}}_{R} \bigr] +{\mbox{\boldmath$F$}}_{\rm rev}+ {\mbox{\boldmath$F$}}_{\rm act}, 
\label{Navier-Stokes}
\end{eqnarray}
where 
\begin{eqnarray}
{{\stackrel{\leftrightarrow}{\mbox{\boldmath$\sigma$}}}} = \eta(\nabla{\mbox{\boldmath$v$}}^\dagger+\nabla{\mbox{\boldmath$v$}}) 
\end{eqnarray}
is the viscous stress tensor, $p$ is the pressure, and ${\stackrel{\leftrightarrow}{\mbox{\boldmath$\delta$}}}$ is the unit tensor. 
The hydrostatic pressure $p$ is determined by the incompressibility condition 
\begin{eqnarray}
\nabla\cdot\mbox{\boldmath$v$}=0.
\end{eqnarray}  
In (\ref{Navier-Stokes}), ${{\stackrel{\leftrightarrow}{\mbox{\boldmath$\sigma$}}}}_{R}$ is the random stress tensor which, in three dimensions, satisfies the fluctuation-dissipation relation as follows  
\begin{eqnarray}
&&\langle\sigma_{R,ij}(\mbox{\boldmath$r$},t)\sigma_{R,i'j'}(\mbox{\boldmath$r$}',t')\rangle\nonumber \\ 
&&=2\eta(\mbox{\boldmath$r$}) T \biggl(\delta_{ii'}\delta_{jj'}+\delta_{ij'}\delta_{ji'}-\frac{2}{3}\delta_{ij}\delta_{i'j'}\biggr)\delta(\mbox{\boldmath$r$}-\mbox{\boldmath$r$}')\delta(t-t'), \nonumber \\ 
\end{eqnarray} 
where $T$ is the temperature in units of the Boltzmann constant. In (\ref{Navier-Stokes}), ${\mbox{\boldmath$F$}}_{\rm rev}({\mbox{\boldmath$r$}})$ is the (volumetric) reversible force density arising from the direct interaction potential between the real particles. With the total potential energy $U$, whose explicit form is provided below, ${\mbox{\boldmath$F$}}_{\rm rev}({\mbox{\boldmath$r$}})$ is uniquely determined as \cite{FPD1,FPD2,FPD3} 
\begin{eqnarray}
{\mbox{\boldmath$F$}}_{\rm rev} = -\sum_{\alpha}\sum_{m={\rm H,T}} \frac{\phi^{\alpha}_m}{\Omega^{\alpha}_m} \frac{\partial U}{\partial \mbox{\boldmath $R$}_{m}^{\alpha}}, \label{reversible_force}
\end{eqnarray}  
where $\Omega_{m}^{\alpha}= \int d{\mbox{\boldmath$r$}}\phi_m^{\alpha}$ is the particle volume. Finally, ${\mbox{\boldmath$F$}}_{\rm act}({\mbox{\boldmath$r$}})$ is the active force density represented as 
\begin{eqnarray}
{\mbox{\boldmath$F$}}_{\rm act} = \sum_{\alpha}\sum_{m={\rm T,P}} \frac{\phi^{\alpha}_m}{\Omega^{\alpha}_m} f_{\rm act}\hat{\mbox{\boldmath$n$}}_\alpha (\delta_{m,{\rm T}}-\delta_{m,{\rm P}}), \label{active_force}
\end{eqnarray}  
where $\delta_{m,l}$ ($m,l={\rm T,P}$) is the Kronecker $\delta$. Notice that the volume integral of ${\mbox{\boldmath$F$}}_{\rm act}({\mbox{\boldmath$r$}})$ is equal to zero, which ensures the conservation of the total momentum. 

In FPD, the velocity of a particle is defined as the following average value: 
\begin{eqnarray}
{\mbox{\boldmath$V$}}_m^{\alpha} =\frac{1}{\Omega_m^{\alpha}}\int d{\mbox{\boldmath$r$}}
\phi_m^{\alpha} {\mbox{\boldmath$v$}}. \label{particle_velocity}
\end{eqnarray} 
It is worth noting that without external forces, active forces, or thermal fluctuations, the time derivative of the total energy of the system, which is the sum of $U$ and the kinetic energy $F_K\{\mbox{\boldmath$v$}\} =\int d\mbox{\boldmath$r$} \rho v^2/2$, is negative definite: ${d}( U+F_K)/dt = -\int d{\mbox{\boldmath$r$}} [({\eta}/{2})\sum_{i,j}(\nabla_iv_j+\nabla_jv_i)^2] \le 0$.

To complete the model of a microswimmer, we must introduce suitable dynamical rules for the phantom particles. We assume that such a particle merely follows the motion of the dimer of real particles to which it is attached. In other words, the position ${\mbox{\boldmath$R$}}_{\rm P}^\alpha(t)$ obeys 
\begin{eqnarray}
{\mbox{\boldmath$R$}}_{\rm P}^\alpha(t) = {\mbox{\boldmath$R$}}_{\rm T}^\alpha(t)-\ell_1 \hat{\mbox{\boldmath$n$}}_{\alpha}(t),  \label{phantom}
\end{eqnarray} 
where $\ell_1$ is the constant center-to-center distance between the tail and phantom particles. We assume that there is no interaction between the phantom particles which therefore can overlap each other. On the other hand, an overlap between the real and phantom particles leads to an unphysical effect: In our model, the real particle is regarded as a rigid particle, and the volumetric force acting on the rigid particle is treated as a homogeneous force density. 
However, any penetration of the phantom particle into the real particle results in an inhomogeneity of the force density within the rigid particle region, creating a contradiction between the required dynamics and the precepts of the FPD numerical scheme.

There are at least two possible ways to avoid any such effect \cite{comment2}. The first one is to introduce an additional dynamical rule, by which the active force is switched off, when the phantom particle of one swimmer overlaps on the body part of the other. The second one is to introduce additional repulsive interactions involving the phantom particles, which directly prevent such overlaps without losing the propulsive force (thus, in this case, the ``phantom'' particle is not exactly a phantom). In this study, we mainly investigate the first of these models, but the second will be discussed in the Appendix with some simulation results presented there for comparison. We find that in the present dumbbell model, the qualitative picture regarding the role of hydrodynamic interactions is relatively insensitive to such model details (although they do alter the quantitative results): the nonlocal and long-range nature of HIs allows them to dominate some important aspects of the collective dynamics. 

Here, we detail the dynamical rule employed in the main text: when $|{\mbox{\boldmath$R$}}_{m}^\alpha - {\mbox{\boldmath$R$}}_{\rm P}^{{\alpha}'}|< a+b$, where $m=({\rm H, T})$ and $\alpha\ne \alpha'$, the $\alpha'$-th swimmer becomes passive. That is, if a swimmer's phantom particle overlaps with the real-particle `body' of another swimmer, the propulsive force pair attached to the first swimmer becomes switched off until its phantom particle once again lies in a purely fluid region. Although somewhat {\it ad hoc}, this dynamical rule could reflect a physical situation in which the motility of a bacterium is weakened if its flagellar bundle encounters the body of a second bacterial body. 
This behavior (a slowing of propulsion in the neighborhood of other particles) will cause an effective attraction between swimming particles \cite{Tailleur_Cates} and will therefore promote clustering. However, such an effective attraction is already present, even for dynamical rules in which the propulsive force is maintained throughout a collision, since the hard-core repulsions still cause swimming particles to slow down at high densities. 
The effects of the rule- and collision-induced slowing are both relevant for the clustering as shown in Sec. III C.
The important aspect for our study is not the local details of our overlap rule but the fact that, at all times, momentum conservation is still satisfied. This will allow us below to make direct comparisons between the chosen dynamics with and without HIs. 

The total potential energy of the system is written as 
\begin{eqnarray}
U\{{\mbox{\boldmath$R$}}_{m}^{\alpha}\} &=& 
\sum_{\alpha} w(|{\mbox{\boldmath$R$}}^{\alpha}_{\rm H}-{\mbox{\boldmath$R$}}^{\alpha}_{\rm T}|) \nonumber \\ 
~~~~~~~~~&&+\frac{1}{2}\sum_{\alpha\ne \alpha'} \sum_{m,m'} u(|{\mbox{\boldmath$R$}}^{\alpha}_{m}-{\mbox{\boldmath$R$}}^{\alpha'}_{m'}|), 
\label{total_energy}
\end{eqnarray}
where $m=${\rm H,T}. The first and second terms correspond to the intra-swimmer and inter-swimmer interactions, respectively, between the real particles. The head and tail particles in the same swimmer are stiffly connected by the following harmonic potential: 
\begin{eqnarray}
w(r)= \dfrac{1}{2}E_1 \biggl(\frac{r}{\ell_0}-1\biggr)^{2}, \label{interactionSD} 
\end{eqnarray}
where $E_1$ is a positive energy constant and $\ell_0$ is the natural length of the dumbbell, which is regarded below as the swimmer size. We assume the following form of $u$: 
\begin{eqnarray} 
u(r) = E_2 \biggl(\frac{2a}{r}\biggr)^{24},  \label{interaction}
\end{eqnarray} 
where $E_2>0$ is introduced to prevent the overlap of real particles on different swimmers.

The numerical calculations are performed as follows. First, the (off-lattice) particle position ${\mbox{\boldmath$R$}}_m^\alpha(t)$ at time $t$ is given. Next, from Eqs.~(\ref{viscosity}) and (\ref{profile}), we obtain the on-lattice fluid velocity field at time $t+\Delta t$ by solving the Navier-Stokes equation, Eq.~(\ref{Navier-Stokes}). Finally, we update the particle position as follows: 
${\mbox{\boldmath$R$}}_m^\alpha(t+\Delta t) = {\mbox{\boldmath$R$}}_m^\alpha(t)+ \Delta t{\mbox{\boldmath$V$}}_m^\alpha(t+\Delta t)$, where ${\mbox{\boldmath$V$}}_m^\alpha(t+\Delta t) = ({1}/{\Omega_m^\alpha})\int d{\mbox{\boldmath$r$}} \phi_m^\alpha {\mbox{\boldmath$v$}}(t+\Delta t)$. 

In implementing our simulations, we first make the equations dimensionless by measuring space and time respectively in units of $\lambda$, which is the discretization mesh size used when solving Eq.(\ref{Navier-Stokes}), and $\tau=\rho\lambda^2/\eta_\ell$. (The latter represents the momentum diffusion time across the unit length.) A mass unit is then chosen to make $\lambda/\tau$, $\bar\sigma = \rho(\lambda^2/\tau^2)$ and $\epsilon=\bar\sigma\lambda^3$ the units of velocity, stress and energy, respectively. 
We set $E_2= 10 \epsilon$ and $E_1= 3.2 \times 10^5 \epsilon$. Because $E_1 \gg E_2$, $\ell_0$ is nearly constant for the particle densities studied here. In addition, the temperature is, unless stated otherwise, set as $T=0.125 E_2$. This value ensures that the effective hard-core particle radius set by Eq.(\ref{interaction}) stays close to $a$ and, in combination with subsequent parameter choices, sets the swimmers' P\'eclet number to a reasonable value (see Sec.\ref{peclet} below). To avoid cumbersome notation we use the above-defined units for the rest of the paper, so that the relevant symbols from now on represent scaled variables.  Throughout our simulations we choose $a_{\rm H,T}=a=3.2$, $a_{\rm P}=0.75a=2.4$, $\xi_{\rm H}=\xi_{\rm T}=\xi_{\rm P}=1$, $\ell_0=2.5a$, and $\ell_1=3a$. The viscosity ratio is $\eta_{\rm p}/\eta_0 =50$. The simulation boxes used are of size $64^3$ and $128^3$.

\section{Simulation Results}
\subsection{Flow field induced by a single swimmer}
\begin{figure}[hbt] 
\includegraphics[width=0.85\textwidth]{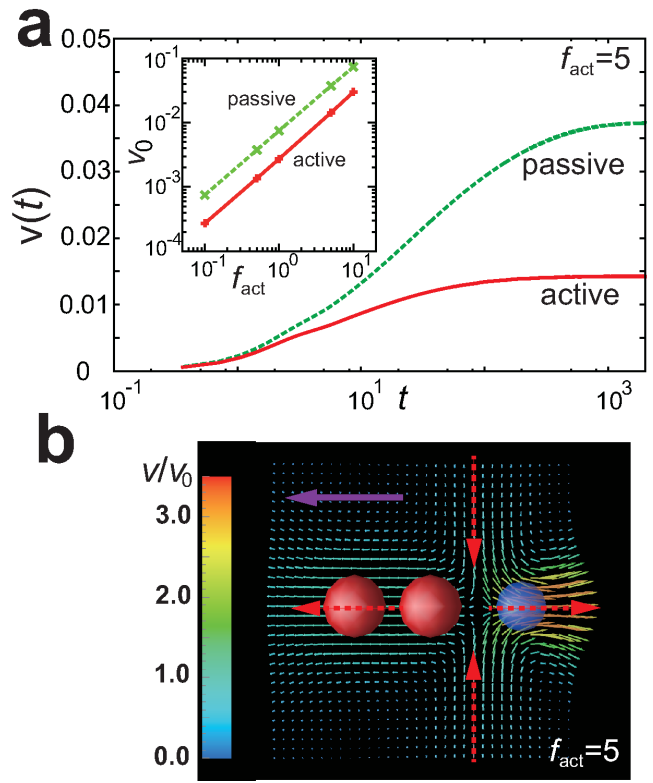}
\caption{(Color online) 
(a) The time evolution of the velocity of an isolated swimmer. Its steady-state value for various values of $f_{\rm act}$ is shown in the inset. The results for the passive dumbbell with an unbalanced external force of the same magnitude are also plotted. 
(b) The velocity field ${\mbox{\boldmath$v$}}(x,y,z=0)$ in the steady state at $f_{\rm act}=5$, where the axis of the swimmer is in the plane. The color bar represents the value of the velocity field scaled by the steady-state swimmer velocity. The purple arrow indicates the swimming direction. The simulation box used is $128^3$ and the frame size shown is $32^2$. }
\label{Fig2} 
\end{figure}

Before proceeding to the many-body results, we investigate hydrodynamic effects in a few-body system, and show that the present model can reproduce the general features of the flow field induced by a single swimmer and by a pair of swimmers. The simulation box used here is $L^3=128^3$. To assist this comparison we set $T=0$ so that the dynamics is deterministic. 
Figure \ref{Fig2}(a) shows the time evolution of the swim speed $v(t)$ when the active force $f_{\rm act}$ is suddenly applied at $t=0$. In the inset of Fig.~\ref{Fig2}(a), the steady-state velocity $v_0$ is plotted against $f_{\rm act}$. Here, we define the Reynolds number of a swimmer as Re=$\rho v_0\ell_0/\eta_0$, which varies from 0.0024 to 0.24 as $f_{\rm act}$ changes from 0.1 to 10. 
These are unrealistically large Reynolds numbers for actual bacteria but, so long as they remain well below unity, the resulting physics should not be strongly affected (see \cite{Cates_Bench} for a fuller discussion in the colloidal context).

The results for a passive dumbbell with an external force $f_{\rm ex}$ of the same magnitude exerted on the tail particle (namely, without the phantom particle) are also shown in Fig.~\ref{Fig2} (a). The active swimmer is found to reach the steady-state velocity faster than does the passive dumbbell. Because of the dipolar nature of the swimmer, the distorted region of the velocity field in the steady state is smaller than that caused by the passive dumbbell, and therefore, the time necessary to reach the steady-state velocity field via momentum diffusion is shorter. 
The scaled friction coefficient along the axis is defined as $\tilde\zeta_{||} \equiv\zeta_{||}/\pi\eta_0\ell_0 = ({\mathcal F}_{\rm act (\rm ex)}/v_0)/\pi \eta_0\ell_0$ for an active (passive) dumbbell, where ${\mathcal F}_{\rm act (\rm ex)}=f_{\rm act(\rm ex)}\int d{\mbox{\boldmath$r$}} \phi_{\rm T}^2/\Omega_{\rm T}=0.59 f_{\rm act(ex)}$ \cite{comment3}.  
We find $\tilde\zeta_{||}\cong 7.8$ for active and $\cong 3.1$ for passive particles.
$\tilde\zeta_{||}$ of the active dumbbell is larger than that of the passive one, which is because the phantom particle, behaving as a model flagellum, pulls the fluid surrounding the dumbbell backwards for $f_{\rm act}>0$.  
Figure \ref{Fig2}(b) shows the velocity field around a swimmer in the steady state for $f_{\rm act}=5$, where the axis of the swimmer  $\hat{\mbox{\boldmath$n$}}$ lies in the plane along the $x$ direction. 

\label{zeta}

In the steady state, the force balances for the head and tail particles are given by
\begin{eqnarray}
{\mathcal F}_{\rm H}^{\rm vis}+{\mathcal F}_{\rm H}^{\rm sp}&=&0,  \\
{\mathcal F}_{\rm T}^{\rm vis}+{\mathcal F}_{\rm T}^{\rm sp}+{\mathcal F}_{\rm T}^{\rm act} &=&0,
\end{eqnarray}
where ${\mathcal F}_{m}^{\rm vis}$, ${\mathcal F}_{m}^{\rm sp}$ ($m=$H,T), and ${\mathcal F}_{\rm T}^{\rm act}$ are the viscous drag, constraining spring, and active forces, respectively \cite{comment1}. 
Because of Newton's third law, ${\mathcal F}_{\rm H}^{\rm sp}+{\mathcal F}_{\rm T}^{\rm sp}=0$, and thus, ${\mathcal F}_{\rm H}^{\rm sp}=[{\mathcal F}_{\rm act}-({\mathcal F}_{\rm H}^{\rm vis}-{\mathcal F}_{\rm T}^{\rm vis})]/2$ and ${\mathcal F}_{\rm T}^{\rm sp}+{\mathcal F}_{\rm act}=[{\mathcal F}_{\rm act}+({\mathcal F}_{\rm H}^{\rm vis}-{\mathcal F}_{\rm T}^{\rm vis})]/2$. Note that these expressions for ${\mathcal F}_{\rm H}^{\rm sp}$ and ${\mathcal F}_{\rm T}^{\rm sp}$ in the steady state do not depend on the method of decomposing the active force into the head and tail particles. In our model the active force acts only on the tail particle, and because of the front-back asymmetry of the streamlines, an imbalance in the hydrodynamic drag between the head and tail particles arises. That is, the hydrodynamic drag acting on the tail particle is stronger than that of the head particle, $|{\mathcal F}_{T}^{\rm vis}|>|{\mathcal F}_{H}^{\rm vis}|$, which is caused by the active force exerted through the phantom particle, resulting in $|{\mathcal F}_{\rm T}^{\rm sp}+{\mathcal F}_{\rm act}|>|{\mathcal F}_{\rm H}^{\rm sp}|$. This is not the case for a passive dumbbell subject to an external force: there, because of the front-back symmetry of the streamlines, the viscous drag forces acting on the head and tail particles are then the same.

\subsection{Flow field induced by a pair of swimmers}

The hydrodynamic pair interactions of swimmers have been extensively studied by many authors in several dumbbell models \cite{Gyrya_Aranson_Berlyand_Karpeev,Pooley_Alexander_Yeomans,Alexander_Yeomans} and in squirmers \cite{Ishikawa_Simmonds_Pedley,Llopis_Pagonabarraga}. 

We here investigate the HIs between two swimmers in several situations relevant to our subsequent results on many-body suspensions. 

\subsubsection{Swimming side by side}
\begin{figure*}[tb] 
\includegraphics[width=0.88\textwidth]{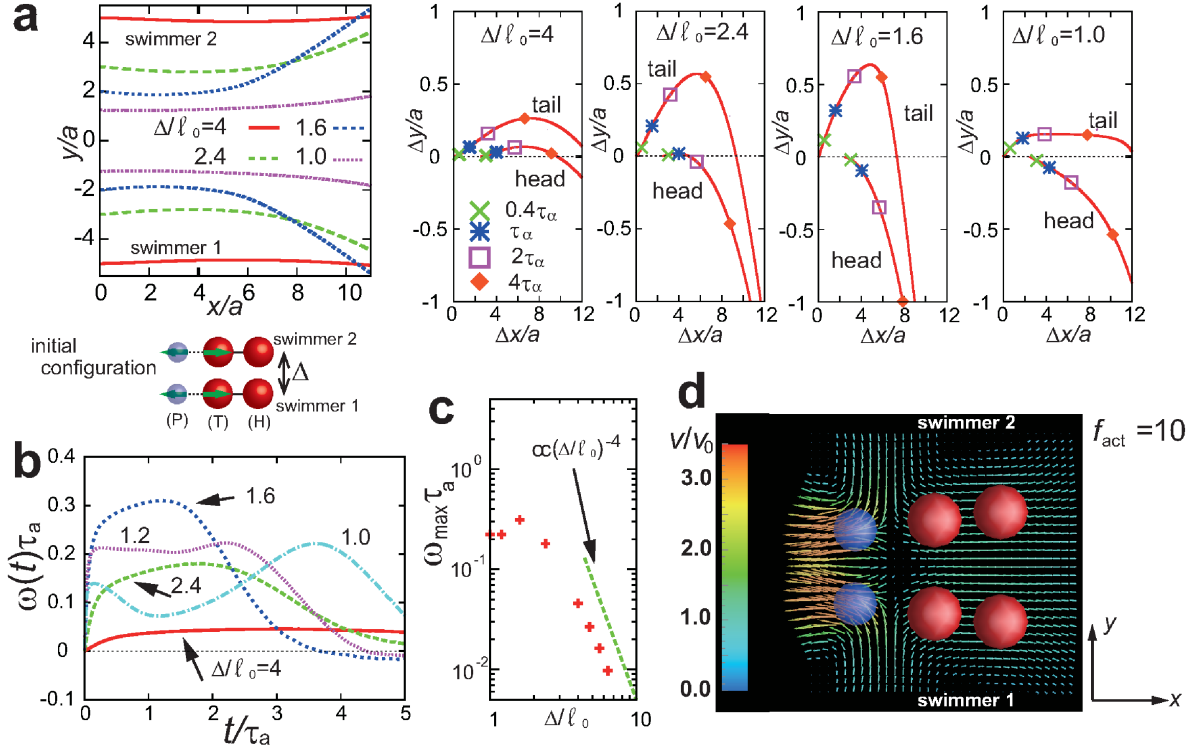}
\caption{(Color online) 
(a) The left-most panel represents the trajectories of the center of mass of the swimmers 1 and 2, where the swimming direction is from left to right and the initial configuration of the two swimmers is shown below this panel. The right four panels represent the trajectories of the head and tail particles of the swimmer 1 for each of the initial separations, where the positions at  several times are explicitly indicated and the ratio of the horizontal scale to the vertical one is about 1/14. Here, the initial position of the tail particle of the swimmer 1 is set to the origin; that is, $\Delta x =x-R_{{\rm T},x}^1(0)$ and $\Delta y =y-R_{{\rm H},y}^1(0)$. In (b) and (c), the time dependence of the angular velocity (angles measured in radians) around the center of mass and its maximum value are shown, respectively.  
For $\Delta/\ell_0\gg 1$, the swimmers hardly change their swimming directions while traveling a distance of their own size ($\sim\ell_0$). However, with decreasing the separation between the swimmers, stronger distortions of the swimming directions are found: In the present case, the hydrodynamic attraction acting on the tail particle is stronger than that on the head particle, and this asymmetry is enhanced for smaller $\Delta$, which leads to faster rotation. 
However, as $\Delta$ decreases further ($\Delta/\ell_0\lesssim 1.6$), this repelling tendency gets weaker, which may be ascribed to the near-field hydrodynamic and steric effects (see the text for discussion). 
(d) Snapshot of two swimmers swimming side by side for $f_{\rm act}=10$ and $\Delta/\ell_0=1.6$ at $t\cong \tau_{\rm a}$. At $t=0$, they are parallel. We also show the velocity field ${\mbox{\boldmath$v$}}(x,y,z=0)$, where the axes of the two swimmers lie in the plane. The color bar represents the magnitude of the fluid velocity scaled by the value of $v_0$ at $f_{\rm act}=10$ shown in the inset of Fig.~\ref{Fig2}(a).}
\label{Fig3}
\end{figure*}

At $t=0$, two swimmers are placed in parallel in the $(z=0)$ plane, and active forces parallel to the $x$-axis are suddenly applied. In Fig.~\ref{Fig3}(a), we plot the trajectories of the head and tail particles of one swimmer for several different initial distances $\Delta$ between the two swimmers. Although both the head and tail particles are initially attracted, they eventually change direction and repel one another. This behavior has been previously observed by several authors \cite{Gyrya_Aranson_Berlyand_Karpeev,Llopis_Pagonabarraga}. A snapshot of the resulting flow field is shown in Fig.~\ref{Fig3}(d). For a symmetric dumbbell swimmer \cite{Graham1,Graham2}, for which the streamlines exhibit front-back symmetry, the hydrodynamic attraction acts equally on both the head and tail particles in the situation considered here. The repulsion we observe is caused by the front-back asymmetry of the streamlines; the hydrodynamic attraction acting on the tail particle is stronger than that on the head particle, and this imbalance results in a torque, causing outward rotation of the swimming direction. 

For $\Delta/\ell_0 \gg 1$ the characteristic rotation time of the swimmer can be estimated using the usual Stokeslet point-force approximation as follows: 
Here, we suppose that the two swimmers are placed in the $xy$ plane and swim in parallel in the $x$ direction. The $y$ component of the velocities of the head and tail particles of swimmer 1 created by swimmer 2 are given by 
\begin{eqnarray}
V_{\rm H,1}^y &=& G_{yx}(\ell_0,\Delta)({\mathcal F}_{\rm T,2}^{\rm sp}+{\mathcal F}_{\rm T,2}^{\rm act})+G_{yx}(\ell_0+\ell_1,\Delta){\mathcal F}_{\rm P,2}^{\rm act}, \nonumber \\ \\
V_{\rm T,1}^y &=& G_{yx}(-\ell_0,\Delta){\mathcal F}_{\rm H,2}^{\rm sp}+G_{yx}(\ell_1,\Delta){\mathcal F}_{\rm P,2}^{\rm act}, 
\end{eqnarray}  
where $G_{yx}(x,y)=xy/8\pi\eta_0 (x^2+y^2)^{3/2}$ is the $yx$-component of the Oseen tensor, and ${\mathcal F}_{m,2}^{\rm sp}$ ($m=$H,T) and ${\mathcal F}_{m,2}^{\rm act}$ ($m=$T,P) are the constraining spring and active forces acting on the swimmer 2, respectively. 
The rotation is given by 
\begin{eqnarray}
\omega= \dfrac{1}{\ell_0}(V_{\rm H,1}^y-V_{\rm T,1}^y), 
\end{eqnarray}
which behaves as $\omega \sim  (v_0/\ell_0) (\ell_0/\Delta)^4  $ for $\Delta/\ell_0 \gg 1$. Here, $v_0\sim f_{\rm act}/\zeta_{||}$ is the swim speed, $\zeta_{||}$ is the friction coefficient along the swimming axis of the dumbbell estimated in the previous subsection, and the relations, $\ell_0=2.5a$ and $\ell_1=3a$, are made use of.  
Thus, for $\Delta/\ell_0\gg 1$, initially parallel swimmers can keep moving in their original directions for a distance much greater than $\ell_0$. However, extrapolating this result to $\Delta/\ell_0\simeq 1$ indicates that during the time $\tau_{\rm a}=\ell_0/v_0$, over which the swimmers travel distances of order their own size, their swimming directions are strongly altered. This is visible in the results for $\Delta/\ell_0=1.6$ shown in Fig.~\ref{Fig3}(a).
 In Figs.~\ref{Fig3}(b) and \ref{Fig3}(c), we show the time dependence of the angular velocity around the center of mass and its maximum value, respectively.  
This tendency is maximized for $\Delta/\ell_0 \simeq 1.6$; down to this value, decreasing $\Delta/\ell_0$ enhances the rotation. 
Our rough estimation of the characteristic rotation $\omega \sim (\Delta/\ell_0)^{-4}$ is qualitatively consistent with the simulation results in this range.

\label{stokeslet}

However, as $\Delta/\ell_0$ decreases further, this repelling tendency gets weaker, which is evident in the last panel ($\Delta/\ell_0$=1.0) of Fig.~\ref{Fig3}(a) and in Figs.~\ref{Fig3}(b) and \ref{Fig3}(c). This happens for two reasons. First, near-field hydrodynamic interactions generically prevent two nearly contacting tail or head particles from changing their separations: this ``squeeze-film'' lubrication drag becomes stronger for smaller $\Delta/\ell_0$. This results in slower rotations, and hence a slower onset of the resulting repulsions. Second, once the tail particles closely approach one another, their direct interparticle repulsion itself prevents further motion. Because of these two effects, when two initially parallel swimmers lie very close to one another, their rotational velocity becomes much smaller than the far-field estimate. As a result they stay close to each other for a time much larger than $\tau_{\rm a}=\ell_0/v_0$. 

\subsubsection{Passing each other}
\begin{figure*}[hbt] 
\includegraphics[width=0.88\textwidth]{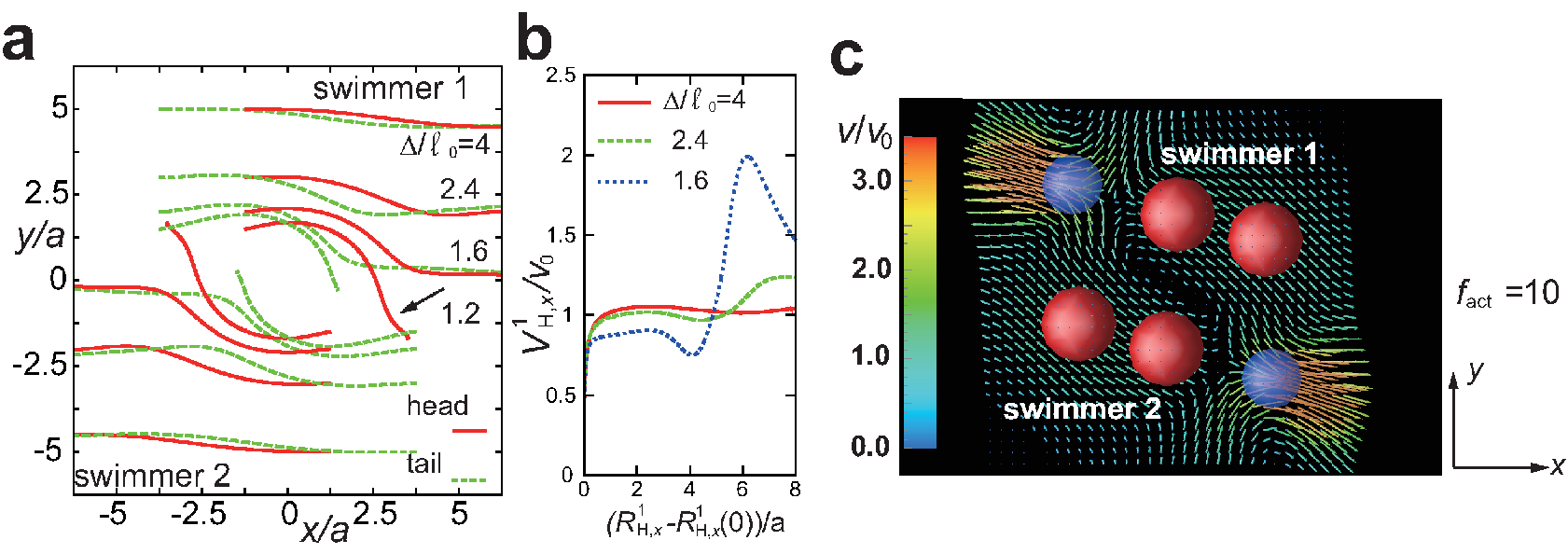}
\caption{(Color online) 
(a) Trajectories of the head and tail particles of the two swimmers for several initial positions. Here, $\Delta=R_{{\rm H},y}^1-R_{{\rm H},y}^2$ at $t=0$. The swimmer 1(2) is swimming from left(right) to right(left). For $\Delta/\ell_0=1.2$, the head particle of each swimmer touches the phantom particle of the other swimmer, and thus, due to the dynamical rule introduced in the present model, the two swimmers stop moving completely. (This would not happen for asymmetric initial conditions, nor in the presence of  thermal fluctuations, or disturbance by the flow caused by other swimmers.) In (b), the velocity of the swimmer 1 along the $x$ axis, $V_{{\rm H}, x}^{1}$, is plotted against $R_{{\rm H}, x}^{1}$. 
 Initially, when the two swimmers are approaching, they repel each other, because the repulsive HIs due to the incompressibility of the solvent are dominant. Then, when they are passing each other, an activity-induced hydrodynamic attraction dominates; that is, one swimmer is attracted to the streamline induced by the other swimmer, resulting in an acceleration of the swim speed. It is evident from the trajectories that such (configuration-dependent) HIs are stronger for smaller separation distances. 
(c) Snapshot of two swimmers passing each other for $\Delta/\ell_0=1.6$ and $f_{\rm act}=10$. We also show the velocity field ${\mbox{\boldmath$v$}}(x,y,z=0)$, where the axes of the two swimmers lie in the plane. The color bar represents the value of the fluid velocity scaled by the value of $v_0$ at $f_{\rm act}=10$ shown in the inset of Fig.~\ref{Fig2}(a).  
}
\label{Fig4}
\end{figure*}

Suppose that at $t=0$ two swimmers lie anti parallel in a staggered coplanar configuration; the active forces are now suddenly applied. Figure~\ref{Fig4}(a) shows the trajectories of the head and tail particles of two such swimmers for several different initial positions. At a large enough separation ($\Delta/\ell_0\gg 1$), the two swimmers pass by each other without a strong distortion of their trajectories. However, with decreasing $\Delta/\ell_0$, the HIs drastically influence their mutual dynamics. When the two swimmers are approaching, they initially repel each other, because the repulsive HIs act on them. Then, when they are passing each other, the head particle of one swimmer is strongly attracted to the streamline induced by the other swimmer, which is also evident in the behavior of the swimming speed shown in Fig.~\ref{Fig4}(b). Therefore at small $\Delta/\ell_0$, the speed of the swimmer first reduces and then accelerates. A snapshot of the swimmers at $\Delta/\ell_0=1.6$ is shown in Fig.~\ref{Fig4}(c). 

Similarly to the analysis presented in Sec. \ref{stokeslet}, by using the far-field Stokeslet point-force approximation, for $\Delta/\ell_0\gg 1$, we can roughly estimate the rotation rate as $(v_0/\ell_0)(\ell_0/\Delta)^4$. The passing time is on the order of $\ell_0/v_0$. Thus, for $\Delta/\ell_0\gg 1$ , the original swimming directions are not much changed during the encounter. On the contrary, an extrapolation of this far-field estimation to a near-field $\Delta/\ell_0\sim 1$ predicts a strong realignment of the swimming directions. This is certainly confirmed by our simulationsas shown in Fig.~\ref{Fig4}. 

\subsection{Dynamics in the semi-dilute regime: Activity-induced clustering } 

So far, we have investigated one and two-body systems and shown that when the distance between two swimmers is comparable to their sizes, the motion of the swimmer is strongly influenced by the flow field caused by the other swimmer. Here, we investigate how such near-field HIs affect the many-body dynamics by simulations in semidilute regimes with thermal fluctuations.  
The simulations presented in this subsection contain $N=320$ swimmers (960 real and phantom particles) in a simulation box with a size of $L^3=128^3$, and thus the volume fraction of real particles is $\Psi=N(\Omega_{\rm H}+\Omega_{\rm T})/L^3=$0.052. In the next section, we will discuss the density dependence of the motility in a smaller simulation box ($L^3=64^3$). In a randomly distributed state, the average distance between the swimmers is approximately $(L^3/N)^{1/3}\sim 2\ell_0$. For comparison we have made equivalent simulations without HIs, using the same parameters for the particle interactions and temperature; without HIs the friction coefficient of a particle is set to $\zeta_{||}/2$ (whose value was evaluated in Sec.~\ref{zeta}). This gives the same value of the swim speed of an isolated swimmer both with and without HIs. 

\label{peclet}
In the presence of thermal fluctuations, the active P{\'e}clet number is defined as Pe$_0 = v_0 \tau_{\rm R}^0/\ell_0$ (so that Pe$_0\sim \zeta_{||} v_0 \ell_0/T \sim f_{\rm act} \ell_0/T$), taking values 8.4 and 16.8 for $f_{\rm act}=7$ and $14$, respectively, as used in the following simulations. Here, $\tau_{\rm R}^0\sim \eta_0\ell_0^3/T$ is the rotational relaxation time of the swimming orientation measured in bulk. Although large Pe$_0$ suggests that active effects dominate over thermal fluctuations, it should be noted that in dilute solution rotational Brownian motion is essential to allow reorientation of the swimming direction.
Another mechanism involves the ``tumbling'' of bacteria at random intervals, causing sudden non-Brownian reorientation; this is not in our model which therefore describes only ``smooth swimming'' bacteria. For these, the chosen P\'eclet numbers are within the achievable range \cite{Bacterial_Parameters}.
Notice also that Pe$_0$ and $v_0$ are values for an isolated swimmer;  in the semidilute regime, due to many-body interactions, a marked slowdown occurs, which will be discussed below.   

The work done by the active force is transformed into the kinetic energy of the solvent and of the swimmers, which is eventually dissipated by viscosity. In the present simulation, the increase of the average kinetic energy density of the solvent due to the active force is at most approximately 1$\%$ of the prescribed value given by the equipartition law,
in which the velocity of a thermal fluid obeys $\langle v_i({{\mbox{\boldmath$r$}}})v_j({{\mbox{\boldmath$r$}}'})\rangle = T\delta_{ij}\delta({{\mbox{\boldmath$r$}}}-{{\mbox{\boldmath$r$}}'})/\rho$ \cite{Landau}, where $i$ and $j$ denote  Cartesian components of ${\mbox{\boldmath$v$}}$. 

\begin{figure}[bth] 
\includegraphics[width=0.45\textwidth]{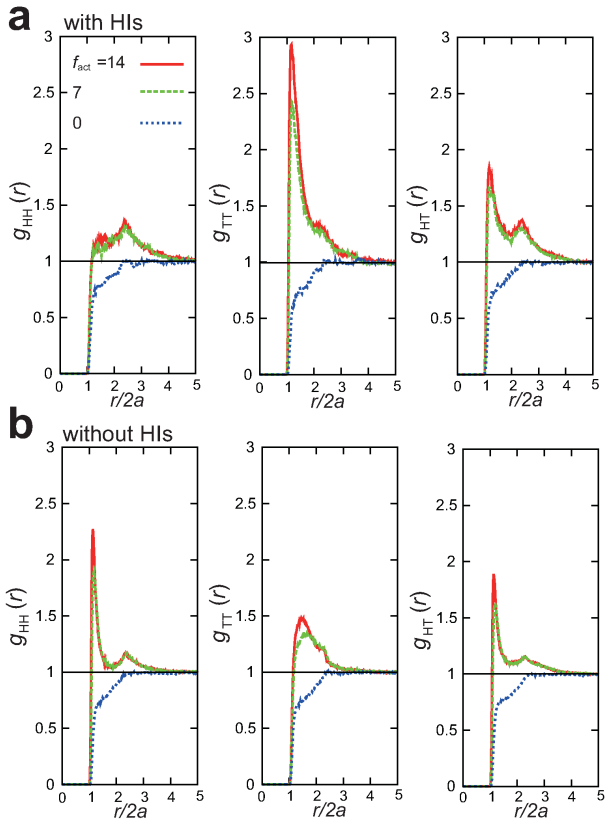}
\caption{(Color online) 
The radial distribution function 
$g_{\rm mm'}(r)$ where $({\rm m,m'})=({\rm H,T})$ with (a) and without (b) HIs. }
\label{Fig5}
\end{figure}
\begin{figure}[hbt] 
\includegraphics[width=0.45\textwidth]{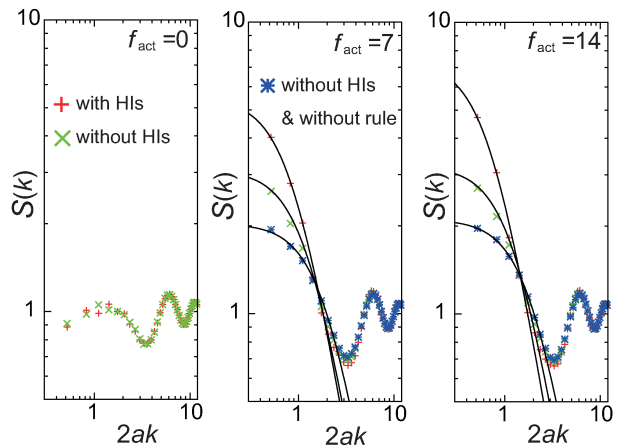}
\caption{(Color online) 
The structure factor $S(k)$ for various values of $f_{\rm act}$. With increasing $f_{\rm act}$, $S(k)$ grows at small $k$. For $f_{\rm act}\ne 0$, $S(k)$ for $2ak\lesssim 1$ can be described by the Ornstein-Zernike form, which is represented by the black solid curves in the cases with and without HIs, respectively.The results with neither the dynamical rule nor HIs are also shown.}
\label{Fig6}
\end{figure}

\begin{figure}[thb] 
\includegraphics[width=0.475\textwidth]{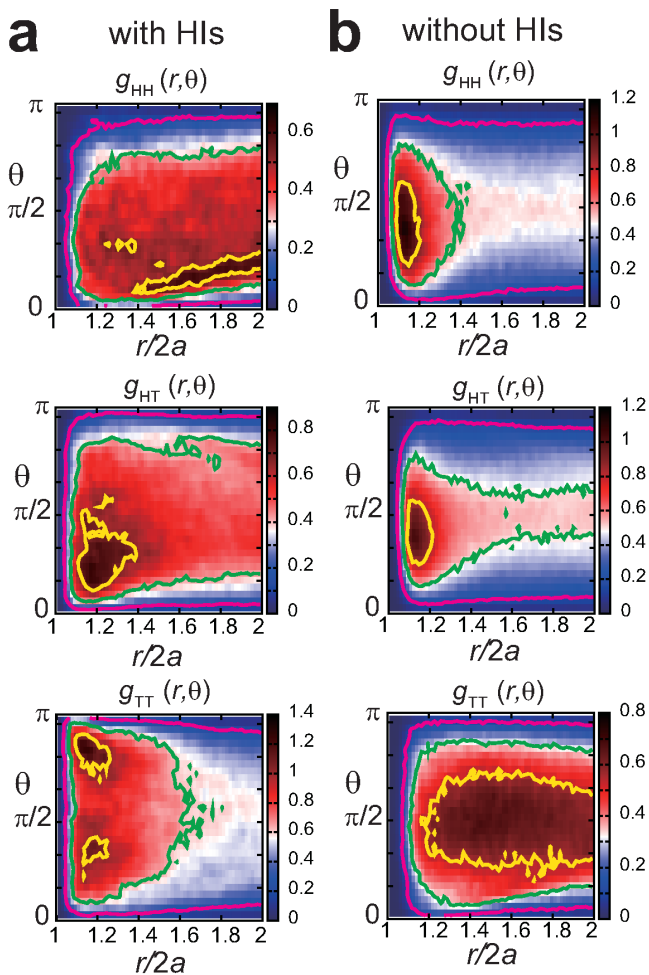}
\caption{(Color online) 
The pair distribution function $g_{mm'}(r,\theta)$ defined as $ g_{mm'}(r,\theta)= \frac{L^3}{4\pi N^2 r^2}\sum_{\alpha\ne \alpha'} \delta[\cos^{-1}({\hat{\mbox{\boldmath$n$}}}_\alpha\cdot{\hat{\mbox{\boldmath$n$}}}_\alpha')-\theta]\delta(|{\mbox{\boldmath$R$}}^{\alpha}_{m}-{\mbox{\boldmath$R$}}^{\alpha'}_{m'}|-r)$ in the steady state at $f_{\rm act} =14$ with (a) and without (b) HIs. The purple, green, and yellow curves are the contours corresponding to 1/6, 3/6, and 5/6 of the maximum value of $g_{mm'}(r,\theta)$, respectively. 
}
\label{Fig7}
\end{figure}
\begin{figure*}[bt] 
\includegraphics[width=1.\textwidth]{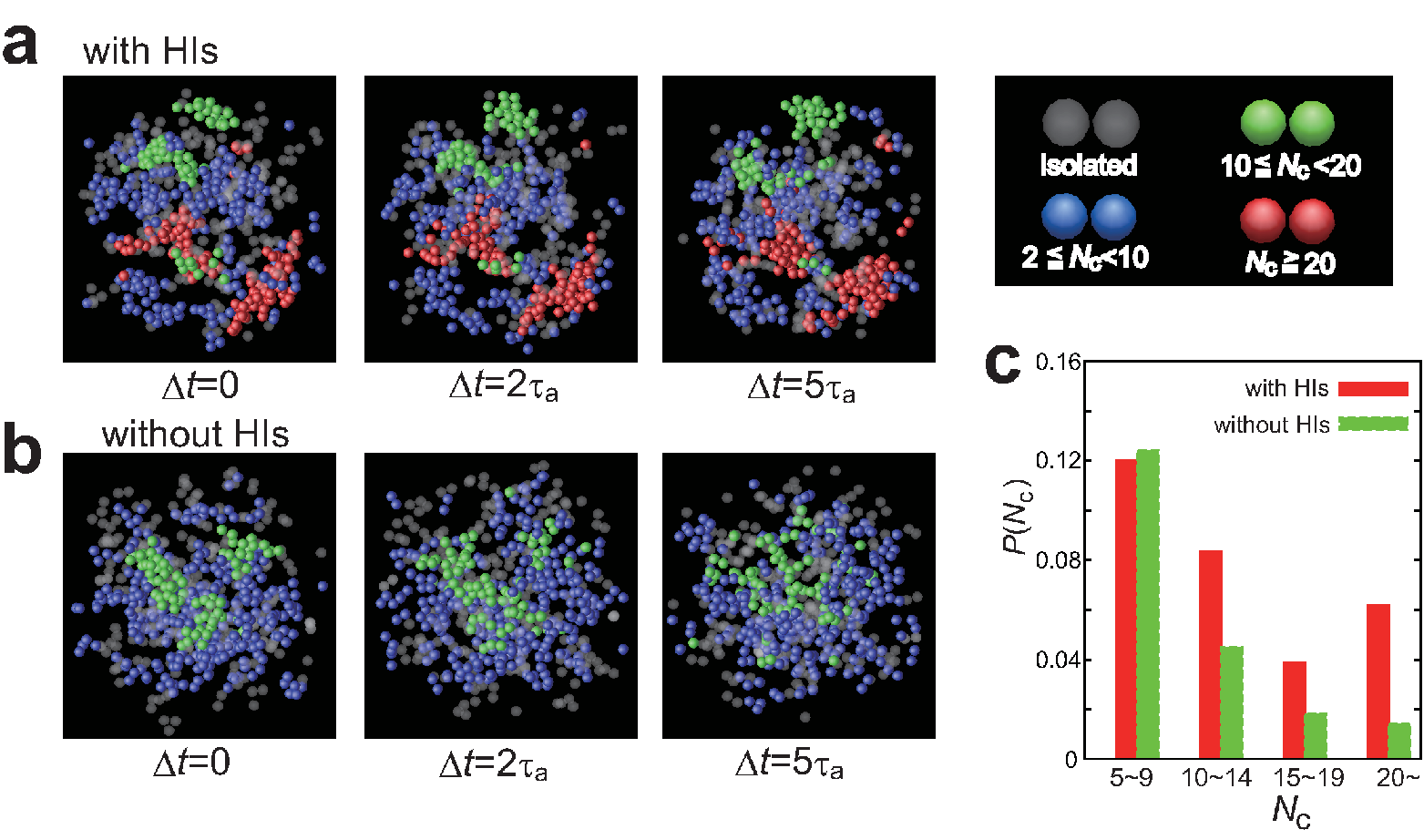}
\caption{(Color online) 
The typical time evolution of the cluster configuration in the steady state at $f_{\rm act}=14$ with (a) and without (b) HIs. Here, colored dumbbells represent the swimmers that compose a cluster at $\Delta t=0$. Different colors represent different numbers of swimmers in a cluster ($N_{\rm c}$). The swimmers which are isolated at $\Delta t=0$ are shown as white semitransparent dumbbells, and the phantom particles are not shown here. The simulation box is $128^3=(40a)^3$ with periodic boundary condition. (c) The probability distribution for a given cluster have $N_{\rm c}$ swimmers with and without HIs. 
}
\label{Fig8}
\end{figure*}

Each simulation starts from a randomly distributed state, and finally reaches a steady state with nontrivial correlations. In Fig.~\ref{Fig5}, we show the radial distribution function 
\begin{eqnarray}
g_{mm'}(r)= \frac{L^3}{4\pi N^2 r^2 }\sum_{\alpha\ne \alpha'}\delta(|{\mbox{\boldmath$R$}}^{\alpha}_{m}-{\mbox{\boldmath$R$}}^{\alpha'}_{m'}|-r) 
\end{eqnarray} 
in the steady state. Both with and without HIs, in an active suspension ($f_{\rm act}\ne 0$), $g_{mm'}(r)$ increases with increasing $f_{\rm act}$, suggesting that the swimmers transiently form clusters. This clustering is more evident from the structure factor given by 
\begin{eqnarray} 
S({{\mbox{\boldmath$k$}}})=1+\dfrac{N}{V}\sum_{m, m'={\rm H,T}}\int d{\mbox{\boldmath$r$}} e^{-i {\mbox{\boldmath$k$}}\cdot{\mbox{\boldmath$r$}}} g_{mm'}({\mbox{\boldmath$r$}}), 
\end{eqnarray}
which is shown in Fig.~\ref{Fig6}. However, there is a marked difference in the behavior with and without HIs; while with HIs the increase is most pronounced in $g_{\rm TT}(r)$, without HIs it is more apparent in $g_{\rm HH}(r)$. More significantly, the overall clustering tendency is significantly enhanced by the addition of HIs. 

These different behaviors should directly reflect the difference in the clustering mechanism with and without HIs. In order to further understand this difference, we consider the following pair distribution function,  
\begin{eqnarray}
&&g_{mm'}(r,\theta)= \frac{L^3}{4\pi N^2 r^2}\nonumber \\ 
&&\times\sum_{\alpha\ne \alpha'} 
\delta[\cos^{-1}({\hat{\mbox{\boldmath$n$}}}_\alpha\cdot{\hat{\mbox{\boldmath$n$}}}_{\alpha'})-\theta]\delta(|{\mbox{\boldmath$R$}}^{\alpha}_{m}-{\mbox{\boldmath$R$}}^{\alpha'}_{m'}|-r), \nonumber\\ 
\end{eqnarray} 
where $\cos^{-1}({\hat{\mbox{\boldmath$n$}}}_\alpha\cdot{\hat{\mbox{\boldmath$n$}}}_{\alpha'})$ is the relative angle between $\alpha$ th and $\alpha'$ th swimmers, and the angular integral of $g_{mm'}(\theta,\psi)$ gives $g_{mm'}(r)$. Figures \ref{Fig7}(a) and \ref{Fig7}(b) represent the steady-state distribution of $g_{mm'}(\theta,\psi)$ around the first peak of $g_{mm'}(r)$ ($1<r/2a<2$), from which we find that neighboring swimmers have different configurations with and without HIs. 

A key role in the clustering in the system without HIs can be ascribed to collision-induced stalling of the swimmers when the head of one particle bumps into another. From Fig. \ref{Fig6}, we find that the present (on/off) dynamical rule further promotes clustering: once swimmers are inactive, the time needed to separate them from their neighbors becomes longer, which promotes further accumulation. Notice, however, that rule-induced inactivity by itself cannot lead to clustering, because Brownian dumbbells ($f=0$) do not cluster without attractive interactions. Clustering (and indeed phase separation) can however result from a density dependence of the mean activity \cite{Tailleur_Cates,Cates_Tailleur}.  Though not a dominant effect at the present density of $\Psi = 0.052$, at higher volume fractions the proportion of the inactive swimmers becomes more significant, resulting in a steep reduction of the motility at what is still relatively modest volume fraction. This point will be discussed in the following section.  
For a perfectly head-to-head collision ($\theta = \pi$) the stalling continues until Brownian motion changes the swimming directions. But such collisions are rare; the dominant effect is from those where the swimming directions are roughly at right angles, giving a larger cross section for collisions but only temporary stalling. The result is a sharp peak of $g_{\rm HH}(r,\theta)$ and $g_{\rm HT}(r,\theta)$ at $r\cong 2a$ and $\theta\sim \pi/2$.  

On the other hand, when HIs are included, these mediate noncontact interparticle forces. As shown previously, these create attractions at large distances, but at intermediate separations cause torques that cause swimmers to separate. At short distances, however, these torques are suppressed and a parallel swimmer in close contact will remain so for a long period. (For antiparallel configurations, there is also a long-lifetime state in which each phantom particle lies on top of the tail particle of the other swimmer, so propulsion is suppressed; unlike for parallel alignment, the existence of this state depends on the chosen propulsion rule.) Such mechanisms of HI-induced trapping should still be effective for the semi-dilute regime studied here. The observed broad peaks in $g_{mm'}(r,\theta)$ at $\theta \lesssim \pi/2$ corresponds to nearly aligned configurations in parallel. On the other hand, the peak in $g_{\rm TT}(r,\theta)$ at $\theta \gtrsim \pi/2$ reflects antiparallel configurations. These results confirm that the activity-induced clustering is driven by different mechanisms depending on whether HIs are present. Without them, particles spend brief but finite periods in close contact due to stalling of rectilinear trajectories; with HIs, they spend longer periods in contact because of trapping in parallel (and, with the chosen swimming rule, also antiparallel) configurations. 

\begin{figure}[bht] 
\includegraphics[width=0.475\textwidth]{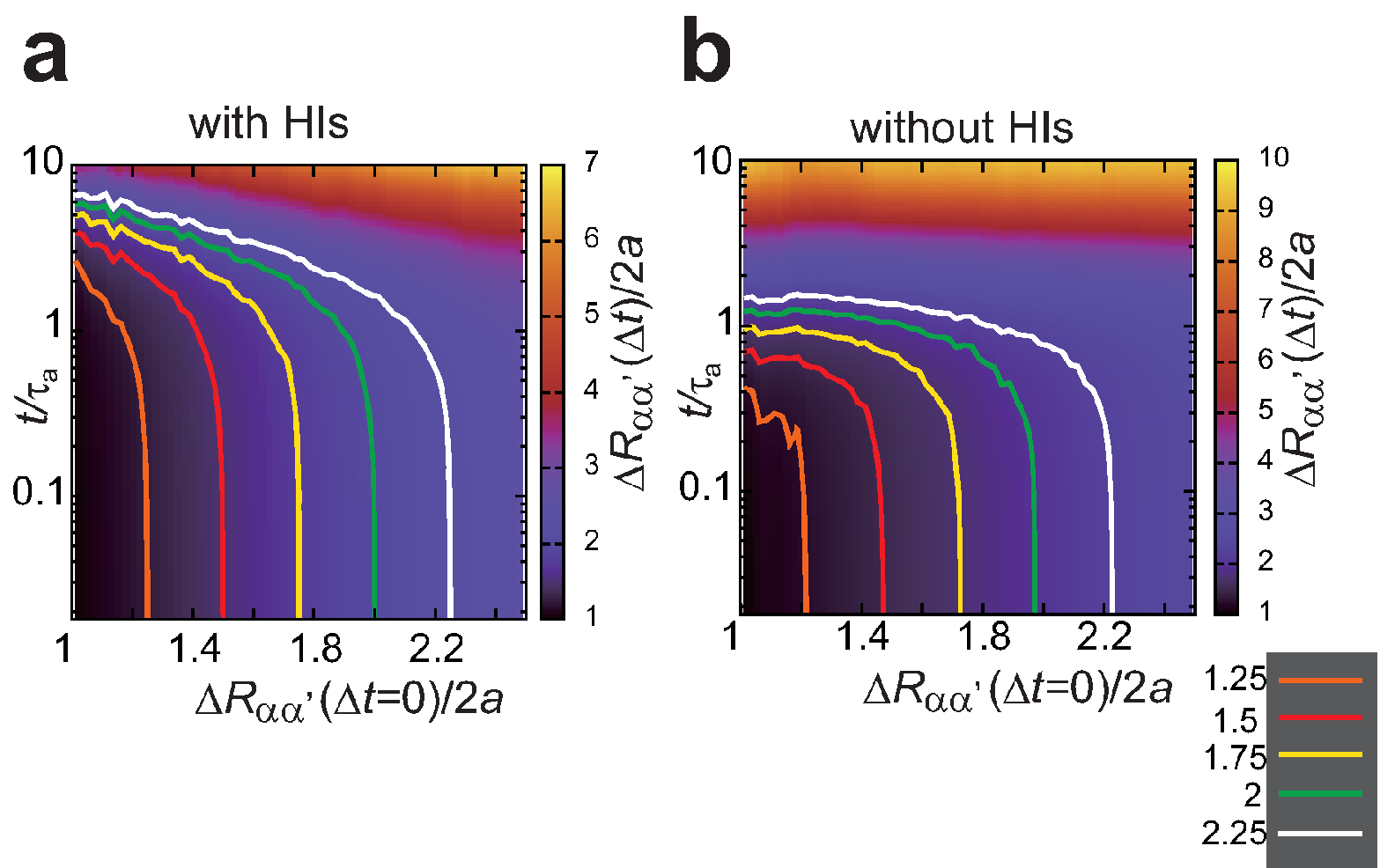}
\caption{(Color online) 
The average separation of a pair of swimmers as a function of its initial value and elapsed time with and without HIs at $f_{\rm act}=14$. Here, the separation is defined as $\Delta R_{\alpha,\alpha'}(t)=|{\mbox{\boldmath$R$}}_{\rm C}^{\alpha}(t)-{\mbox{\boldmath$R$}}_{\rm C}^{\alpha '}(t)|$, where ${\mbox{\boldmath$R$}}_{\rm C}^{\alpha}=({\mbox{\boldmath$R$}}_{\rm H}^{\alpha}+{\mbox{\boldmath$R$}}_{\rm T}^{\alpha})/2$ is the center of mass of the $\alpha$ th swimmer. The colors represent the scaled value of $\Delta R_{\alpha,\alpha'}(t)$.  
With HIs, the initially adjacent pairs $[\Delta R_{\alpha,\alpha'}(\Delta t =0)\lesssim \ell_0(=2.5a)]$ stay close to each other for a longer time than $\tau_{\rm a}$, although without HIs, such swimmers separate themselves much quicker. }
\label{Fig9}
\end{figure}

This difference in the clustering mechanism is responsible for the contrasting steady-state properties of clusters, such as their lifetime and the average size, which are directly visible from the sequential simulation snapshots of the swimmers as shown in Fig.~\ref{Fig8}.  
When identifying clusters, the $\alpha$ th and $\alpha'$ th swimmers are considered to be connected if 
\begin{eqnarray}
|{\mbox{\boldmath$R$}}_m^\alpha-{\mbox{\boldmath$R$}}_{m'}^{\alpha '}|\le \delta R, 
\end{eqnarray}
for at least one combination of $m$ and $m'$, where we set $\delta R=0.6a$ which is approximately the first-peak width of the radial distribution function. This choice of $\delta R$ is rather arbitrary, but a small change of its value does not essentially change our conclusions. 
In Figs.~\ref{Fig8}(a) and \ref{Fig8}(b), we show  the typical time evolution of clusters with and without HIs, respectively. 
Without HIs transient clusters form due to the mixed effects of the collision- and rule-induced stalling, but with the present parameters, the persistence time of contacts is approximately the time $\tau_{\rm a}=\ell_0/v_0$ for an isolated swimmer to move its own length. Clusters therefore remain weak and transient. On the other hand, in the system with HIs, once the swimmers are close enough they remain in bound-state configurations for a time much longer than ($\tau_{\rm a}$), resulting in stronger and longer-lived clusters. 

We emphasize that many-body hydrodynamic effects further reduce the average swimming speed of our dumbbells, defined as the instantaneous projection of their velocity onto the swimming direction. In our semidilute system ($\Psi=0.052$), the average swim speed is evaluated as $0.41v_0$ and $0.78v_0$ with and without HIs, respectively. 
Without HIs, about 20$\%$ of the swimmers are inactive on average; in this case the reduction of the swim speed is largely owing to the dynamical rule. However, a further large reduction is induced by HIs: with these nonlocal effects switched on, about 35$\%$ of the swimmers are inactive on average. 
This difference in the slowing-down is also seen in Fig.~\ref{Fig9}, where the average time evolution of the separation distance of a pair of swimmers is plotted for various initial values. This clearly shows that with HIs,  prolonged trapping of swimmers becomes more marked than without HIs. 
This hydrodynamic trapping stimulates a further accumulation of the swimmers and subsequently leads to an enhancement of the cluster growth, which is evident in Fig.~\ref{Fig8}(c), where we show the probability distribution for numbers in a cluster with and without HIs. 
Although clusters are longer-lived with HIs, they do finally break up: recall that the hydrodynamic torque tends to misalign the swimming directions and eventually a cluster will find a configuration where this effect dominates long enough to cause its disintegration.

The tendency of HIs to promote clustering is strongly linked to the near-field hydrodynamics, especially in its effects on rotation of the swimming velocity. These effects can be expected to vary with particle shape and swimming type, and indeed our observations are almost the opposite of what happens in (neutral) squirmers \cite{Fielding}, where HIs act to suppress rather than enhance the formation of density inhomogeneities.

\section{Density dependence of the motility}

\label{motility}

In the previous section we showed that, for the swimmers at the chosen volume fraction $\Psi=0.052$, there is a distinct difference in the swimmers' dynamics, and especially their clustering mechanism, with and without HIs. 
As the volume fraction increases, the effects of the many-body interactions (and also of the dynamical rule) on the dynamics should be more pronounced, 
and thus the dynamics is expected to be significantly slower at high densities. Here, we investigate the volume fraction dependence of the mean motility variables (swim speed and rotational relaxation time) and discuss the influence of these dependences on the clustering process. We use the same simulation parameters as in Sec.~\ref{peclet} except for the box size, which is here $L^3=(64)^3$. Note that we checked that the essential results for $\Psi=0.052$ shown in Sec.~\ref{peclet} are barely affected by this change. 

A key observation on increasing the volume fraction far beyond $0.052$ is that the chosen dynamical rule (switching off activity  on overlap of phantom and real particles) has an increasingly pronounced effect at high densities. This motivated the simulations on an alternative model (described in the Appendix) to clarify the role of HIs on the collective behavior. The comparison of these approaches suggests that in the present dumbbell model the details of the local interactions do not qualitatively alter the physical role played by the HIs. For while there are significant near-field hydrodynamic effects which the local interactions clearly do alter, the main role of HIs arises at intermediate and large distances where they can directly affect the collective dynamics. 

\begin{figure}[h] 
\includegraphics[width=0.475\textwidth]{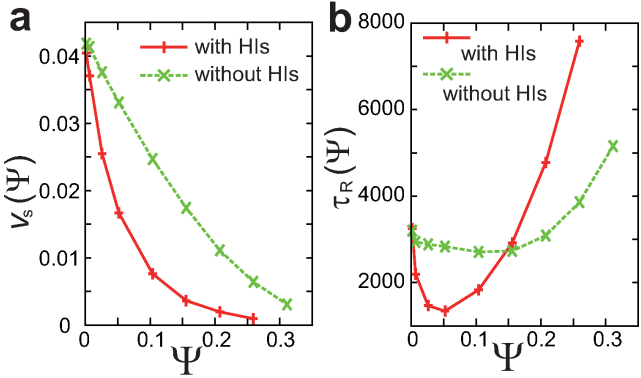} 
\caption{(Color online) 
The average swim speed (a) and the rotational relaxation time (b) for various volume fraction $\Psi$ at $f_{\rm act}=14$. The red and green curves are for the cases with and without HIs, respectively.
}
\label{Fig10}
\end{figure}
 
In Fig.~\ref{Fig10}(a), we plot the dependence on the volume fraction $\Psi$ of the average swim speed, which is here defined by 
\begin{eqnarray} 
v_{\rm s}(\Psi)= \dfrac{1}{2N} {{\sum}_{\alpha}} \langle \hat{\mbox{\boldmath$n$}}_\alpha \cdot({\mbox{\boldmath$V$}}^\alpha_{\rm H} + {\mbox{\boldmath$V$}}^\alpha_{\rm T}) \rangle, 
\end{eqnarray}
where $({\mbox{\boldmath$V$}}^\alpha_{\rm H} + {\mbox{\boldmath$V$}}^\alpha_{\rm T})/2$ is the velocity of the center of mass of the $\alpha$ th swimmer, $\langle \cdots \rangle$ represents the time average in the steady state. 
Without HIs, upon increasing $\Psi$ to moderate values, $v_{\rm s}(\Psi)$ decreases. This is largely because of the chosen dynamical rule, which causes an increase in the number of inactive swimmers at high density: at $\Psi = 0.31$, nearly 90$\%$ of phantom particles overlap with the real particles of other swimmers and thus become inactivated. (Note that the details of this decrease will depend on the swimmer's shape.)
On the other hand, with HIs, $v_{\rm s}(\Psi)$ drops sharply even at rather small $\Psi$, in a regime where there are much fewer inactivation events due to overlap. This reflects the long-range character of HIs, which allow each swimmer's motion to be strongly influenced by that of its rather distant neighbors, promoting collective motions even without direct collisions. Such many-body hydrodynamic effects seemingly increase the mean drag force, resulting in the steep decrease of $v_{\rm s}(\Psi)$. 

A strong effect of HIs is also seen in the rotational dynamics, as characterized by the relaxation time of the orientational correlator   
\begin{eqnarray}
H(t)=\langle \hat{\mbox{\boldmath$n$}}_\alpha(t)\cdot \hat{\mbox{\boldmath$n$}}_\alpha(0)\rangle, 
\end{eqnarray}
which can be fit to an exponential form in the whole range of $\Psi$ studied here. In Fig.~\ref{Fig10}(b), the $\Psi$-dependent rotational relaxation time, $\tau_{\rm R}(\Psi)$, is shown. Without HIs, $\tau_R$ is nearly constant for low $\Psi(\lesssim 0.2)$, but increases to exceed the Brownian relaxation time $[\tau_{\rm R}(\Psi\cong 0)]$ at higher $\Psi(\gtrsim 0.2)$. 
This increase is because steric hindrance to rotation increases with crowding at higher $\Psi$; even in equilibrium ($f_{\rm act}=0$), deviations from the ideal-gas behavior due to the excluded volume effects become pronounced for $\Psi\gtrsim0.2$ (not shown here). On the other hand, upon adding HIs, a striking change in the behavior of $\tau_{\rm R}(\Psi)$ is found: first a sharp decrease as $\Psi$ rises towards about 0.05, and then a clear upturn at higher $\Psi$. 

Both features can be ascribed to the cooperative nature of HIs. 
As was discussed in the previous section, HIs among the swimmers generate torques. At small $\Psi(\lesssim 0.05)$, such hydrodynamic torques should enhance rotational relaxation which would otherwise rely solely on Brownian motion. This extra rotation results in the sharply decreasing  trend in $\tau_{\rm R}(\Psi)$ for small $\Psi$. (Notice that this effect depends on both the rodlike shape and the front-back asymmetric streamlines around the swimmer within the present model.) However, with crowding, the mean separation between swimmers decreases, which should suppress reorientation while enhancing the hydrodynamic attractions. This presumably causes the reversal of the decreasing trend in $\tau_R$ at around $\Psi\sim 0.05$. Increasing the volume fraction further ($\Psi\gtrsim 0.15$), one expects activity-induced flow fields to increasingly be suppressed, reducing motility and increasing $\tau_R$. Nonetheless, due to the incompressible nature of HIs, neighboring swimmers push and pull one another even without direct contact, causing the translational and rotational dynamics to become slower than in the system without HIs. 

\begin{figure}[hbt] 
\includegraphics[width=0.475\textwidth]{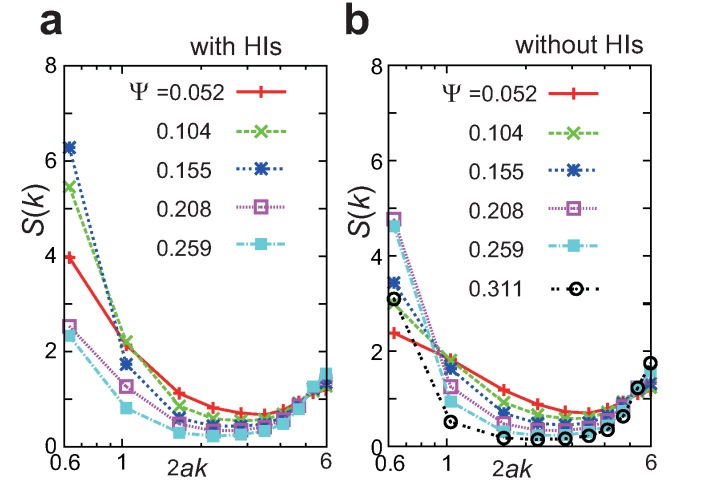} 
\caption{(Color online) 
The structure factor with (a) and without (b) HIs for various $\Psi$ at $f_{\rm act}=14$. 
}
\label{Fig11}
\end{figure}

These differences in the volume fraction dependence of the motility with and without HIs affect the physics of clustering. 
In Fig. \ref{Fig11}, we show the structure factor for various $\Psi$ with and without HIs. Without HIs, the clustering tendency increases with volume fraction until $\Psi\sim 0.25$. On the other hand, with HIs, this increase continues only until $\Psi \sim 0.15$. These clustering tendencies were also checked by direct visualization. 

In recent theories for active Brownian particles (ABPs), it was predicted that, if the swim speed decreases steeply enough with density, local densification induces a further accumulation of particles, eventually leading to unstable growth of fluctuations and phase separation \cite{Tailleur_Cates,Cates_Tailleur,CatesR}. This mechanism for activity-induced phase separation was confirmed by subsequent simulations on self-propelled spherical particles with rotational diffusion and no HIs \cite{Stenhammar_Tiribocchi_Allen_Marrenduzo_Cates}. Although the present dumbbell swimmers can be categorized as ABPs, we have shown that the physics of mutual slowing down depends on particle shape, so that even without HIs there is no exact relationship to these earlier simulations. However, the underlying theoretical analysis has some quite generic features \cite{Tailleur_Cates,Cates_Tailleur,CatesR} so it is useful to discuss the clustering behavior observed here in similar terms.

As shown in Fig.~\ref{Fig10}, without HIs and for $\Psi\lesssim 0.2$,  $v_{\rm s}(\Psi)$ decreases almost linearly with $\Psi$, while the rotational relaxation time $\tau_{\rm R}(\Psi)$ decreases only slightly from the value set by purely Brownian rotation. 
Within the framework of the theory \cite{Tailleur_Cates,Cates_Tailleur,CatesR}, the effective compressibility $S(0)$ should increase with $\Psi$, diverging at a spinodal set by the condition $dv_{\rm s}/d\Psi<- v_{\rm s}/\Psi$. 
This condition holds for $\Psi\gtrsim 0.15$ [evaluated from the fitting of $v_{\rm s}(\Psi)$ in Fig.~\ref{Fig10}]. In this evaluation, $dv_{\rm s}/d\Psi+ v_{\rm s}/\Psi$ is found to have a minimum ($<0$) at $\Psi \cong 0.23$.
The compressibility increase is certainly seen for $\Psi\lesssim 0.25$ as shown in Fig.~\ref{Fig11}(b), but beyond this point, rather than diverging, $S(0)$ falls again even though the spinodal condition is satisfied by the observed $v_{\rm s}(\Psi)$. This suggests that for the dumbbell system at this density, direct repulsion via excluded volume interactions (which are omitted from the theory) are strong enough to overcome the effective attraction caused by the density-dependent motility. However, we have performed additional exploratory simulations without HIs (in a simulation box of $L^3=128^3$) using a larger self-propulsion force; these suggest that at $\Psi=0.2$ and $0.25$ phase separation does occur at high enough values of $f_{\rm act}$. 

A broadly similar scenario holds in the presence of HIs, as seen in Fig.~\ref{Fig11}(a). However the maximum compressibility is significantly greater, and furthermore arises at significantly lower density ($\Psi \simeq 0.15$). 
The peak coincides with the state of most pronounced clustering as observed by direct visualization. The data for $v_{\rm s}(\Psi)$ reported in Fig.~\ref{Fig11}(a) again suggest that the spinodal condition $dv_{\rm s}/d\Psi<- v_{\rm s}/\Psi$ of \cite{Tailleur_Cates,Cates_Tailleur,CatesR} is satisfied for $\Psi\gtrsim 0.06$, yet once more we observe a maximum, not a divergence, of $S(0)$ near the state of maximal clustering at $\Psi\cong 0.15$. [Note that $dv_{\rm s}/d\Psi+v_{\rm s}/\Psi$ now has a minimum ($<0$) at $\Psi\cong 0.12$; above this density, the activity-induced attraction itself becomes weakened.]
The subsequent decrease might be attributable to steric repulsions, but their effects would have to be stronger than in the case without HIs to start dominating at this somewhat lower density. An additional factor could be the effect of HIs on reducing $\tau_R$, which is the mechanism for avoidance of phase separation discussed by Fielding \cite{Fielding}. We have performed additional exploratory simulations with HIs [at the density $\Psi=0.15$ in a simulation box of $L^3=(128)^3$] using a larger self-propulsion force; the structure factor $S(k)$ at small $k$ is found to be dramatically enhanced by formation of a space-spanning structure of the order of the system size, which suggests that bulk phase separation occurs for large enough $f_{\rm act}$. Notice, however, that the formed structure is not compact and exhibits large fluctuations, which may be due to HIs and the relatively small system size of the present simulation. To firmly assess whether bulk phase separation occurs, we need to perform simulations with larger system sizes.

Thus we have shown the observed clustering tendency of our dumbbells to be broadly consistent with a theory in which collisional interactions are represented at mean-field level by a density-dependent swim speed \cite{Tailleur_Cates,Cates_Tailleur,CatesR}. However, using the numerically obtained $v_{\rm s}(\Psi)$ shown in Fig.~\ref{Fig10}, this theory predicts bulk phase separation in a regime where we see only clusters. (This applies both with and without HIs.) Although steric repulsions offer an obvious candidate mechanism for this discrepancy, the detailed conditions for bulk phase separation in the present dumbbell model are not well understood, and a further systematic study will be needed to clarify them. 

\section{Summary and Discussion}

In this study, we have numerically investigated the effects of HIs on the collective dynamics of active suspensions, modeling each micro-organism as a stroke-averaged dumbbell swimmer with a prescribed force dipole (and an overlap rule). Our results can be summarized as follows. 

With HIs, when the separation distance between the swimmers is comparable to their sizes, their swimming motions are strongly influenced by one another. The activity-induced flow field creates both attractive and repulsive effects that depend on the relative positions and orientations of the swimmers. These effects are significantly more complex than could be expected for simpler models of spherical particles (such as squirmers) in which the nonhydrodynamic forces between particles are spherically symmetric. Although in the far field swimming dumbbells attract, at intermediate separations the HI-induced torques tend to disalign the swimming directions causing swimmers to move apart. Then, at closer distances, these repulsive effects are suppressed; hydrodynamic forces promote preferred structures, in which the swimmers are close to parallel alignment (or, with our overlap rule, antiparallel). These states can last much longer than the intrinsic characteristic time for an encounter, $\tau_{\rm a}= \ell_0/v_0$.  
Our simulation results strongly suggest that such activity-induced HIs should be significant at rather smaller volume fraction ($\Psi \sim 5\%$), where they cause strong enhancement of a clustering tendency that is already apparent, albeit in a weaker form and by a different mechanism, when HIs are absent. 

Our simulation results for the density dependence of motility parameters show distinctly different behaviors with and without HIs. 
At a relatively small volume fraction, with HIs, the translational swimming motion becomes slower, while rotational diffusion becomes faster; these are natural consequences of the far-field and intermediate-range pair interactions just described, and are stronger than the corresponding effects without HIs. On the other hand, at large enough volume fractions, the translational and rotational motions are both strongly suppressed. With crowding, because less space is available around a swimmer, the ability of each force dipole to set up a coherent propulsive flow field is diminished, weakening both the motility and the HIs. However, due to the incompressible nature of HIs, neighboring swimmers push and pull each other around even without direct contact. This impedes both the translational and rotational motions much more than is the case without HIs, where only direct collisional forces act between the swimmers. 

The different volume fraction dependences of the motility with and without HIs  influence the tendency to form clusters. For our chosen interaction and propulsion parameters, clustering is maximal at an intermediate volume fraction in each case, but this volume fraction is lower, and clustering stronger, with HIs than without. The conditions under which bulk phase separation arises, rather than just enhanced density fluctuations, remain somewhat unclear; in particular steric or other interactions at volume fractions above $\Psi \simeq 0.15$ (with HIs) or $\Psi\simeq 0.2$ (without) apparently suppress phase separation under conditions where the observed dependence of the mean swim speed $v_{\rm s}$ on $\Psi$ might lead one to expect it \cite{Tailleur_Cates,Cates_Tailleur,CatesR}. Such corrections to the theory have also been reported for spherical self-propelled particles \cite{Stenhammar_Tiribocchi_Allen_Marrenduzo_Cates}, but not at such low densities. 

In considering the effects of HIs on collective dynamics, we note in addition important differences between alternative types of models. In a recent simulation of two-dimensional squirming disks \cite{Fielding}, it was found that the activity-induced phase separation is strongly suppressed by HIs. 
In that system, because the slip velocity is prescribed at their surfaces, when two squirmers come very close, they generally tend to strongly rotate and repel. Thus, crowding enhances the reorientation of the swimmers and reduces their tendency to trap one another. In contrast, in our model of force-prescribed dumbbells, although the reorientation is enhanced at low enough volume fractions, at higher ones  both the translational and rotational motions become much slower than without HIs. Moreover, at a certain intermediate volume fraction, the hydrodynamic trapping effect (and indeed the clustering) is most pronounced. Further simulation studies and analysis will be required to clarify the detailed conditions for phase separation with and without HIs in our dumbbell model.

In previous works by Graham and co-workers \cite{Graham1,Graham2,Graham3}, it was found that, in suspensions of force-prescribed dumbbell pushers, HIs among them cause coherent fluid motions with a correlation length larger than the swimmer size. Interestingly, such correlated motions are hardly seen in suspensions of pullers. Saintillan and Shelley \cite{Saintillan_Shelley1,Saintillan_Shelley2,Saintillan_Shelley3} reported similar results by simulating a similar model, where the active particle is modeled as a stress-prescribed rigid rod. Furthermore, they also found that an isotropic or aligned homogeneous state becomes unstable due to HIs and some non-linear hydrodynamic effects lead to steady-state structure with a pronounced density inhomogeneity and local orientational order of swimmers \cite{comment4}. Although these previous studies do not contradict our simulations, we found that, at certain intermediate densities, the clustering caused by activity-induced attractions (or hydrodynamic trapping) is greater than previously reported, which, as already discussed, can be attributed to the effects of near-field HIs. The advantage of the hybrid simulation methods such as that used in the present study is that, by directly taking the solvent dynamics into account, we can accurately treat HIs among finite-sized particles at moderate separations, and also capture some significant near-field effects.
In the previous studies, on the other hand, active particles are essentially represented as pairs of point forces and HIs are simply evaluated by using the usual Stokeslet. However, such an approach based on far-field hydrodynamics is not able to fully capture the collective hydrodynamic effects in relatively dense active suspensions. Thus, our present study should complement the earlier works cited above.

Our dumbbell simulations include thermal fluctuations, which act like a repulsion to prevent the swimmers from clustering. 
Similar to an analysis performed in Sec.~\ref{stokeslet}, we can use the Stokeslet point force approximation to estimate the energy scale of the dipolar hydrodynamic forces as $f_{\rm act}\ell_0/10$, where the interaction range is assumed to be from the core size ($=2a$) to the mean separation distance of the swimmers ($\sim 2\ell_0=5a$). 
Thus the ratio of this energy scale to the temperature is of order Pe$_0/10$ where Pe$_0$ is the P{\'e}clet number introduced previously. By this rough estimate, thermal and hydrodynamic effects can compete in our simulations. (We have done further exploratory simulations without thermal fluctuations, and found that the clustering is certainly enhanced when these are absent.) 

Some quantitative aspects of our results depend on the nature of the local interactions between swimming dumbbells, in particular the rule that causes propulsion to be switched off when swimmers overlap. However, by varying this rule (see the Appendix) we have found a robust role for long-range hydrodynamic interactions in controlling the collective behavior. The HIs promote clustering through a collective slowing of propulsive motions at densities that are too low for the rules governing direct interparticle collisions to be their dominant cause. In contrast, without HIs the slowing is collisional and thus depends more strongly on the local interactions. Our finding that HIs have significant effects on clustering physics is far from obvious; due to the dipole-dipole nature of these interactions among self-propelled particles, they are much weaker than for particles subjected to external forces, and under many conditions their effects appear almost negligible \cite{Drescher}. Certainly one can argue that for situations where the main interaction between particles involves well-separated two-body collisions, the effects of HIs are indeed relatively weak; however this argument is dangerous when applied to the many-body collective behavior of swimmers at the fairly high densities actually arising within clusters. 

We conclude with some further remarks on topics that we hope to study in future:
 
(i) Collective hydrodynamic effects should strongly depend not only on the shapes of the swimmers but on their  self-propulsion mechanism. 
We have done preliminary studies for pullers ($f_{\rm act}<0$); these show quite different steady-state properties from the pushers studied in this paper. At an equal volume fraction of $\Psi=0.052$, the  near-field radial distribution functions (for separations $r\lesssim \ell_0$) are significantly smaller than those of the equilibrium ($f_{\rm act}=0$) dumbbells. 
This indicates that for pullers repulsive interactions dominate at these separations, in contrast to the case of pushers. 
Accordingly, a very different density dependence of the motility (and therefore different clustering and/or phase separation properties) from that of pushers is expected to arise, whose details remain to be explored.

(ii) In recent experiments on {\em E. coli} in the presence of additional attractive forces (created via a depletion potential due to polymer additives) it was shown experimentally and by simulation that activity produces a significant shift of the phase boundary compared to that of a passivated system with the same attractions \cite{Shwarz_etal}. However the configurations favored by such an attraction need not coincide with those stabilized by the activity-induced HIs. Therefore the case of active dumbbells with attraction requires separate investigation. One very recent study suggests a mechanism whereby the equilibrium phase separation caused by attractions is interrupted by activity-induced cluster breakup \cite{Chantal_Frenkel}, but equilibrium attractions could equally interfere with the motility-induced phase separation mechanism discussed in Section~\ref{motility}.  

{\em Acknowledgments:}
This work was financially supported by the Foundation for the Promotion of Industrial Science, a grant-in-aid for Scientific Research (C) and Specially Promoted Research from JSPS, and EPSRC Grant EP/J007404. MEC holds a Royal Society Research Professorship.
Simulations were partially performed at the the Supercomputer Center, the Institute for Solid State Physics, the University of Tokyo and at ECDF at the University of Edinburgh.  

\appendix
\section{}

In the main text, we have avoided overlap between the propulsive element of one swimmer, represented as a phantom particle, and the bodies of others by introducing a dynamical switch-off rule for the propulsion. In this Appendix, a different model is examined;  instead of using the dynamical rule, we introduce repulsive interactions involving the phantom particles to directly prevent such overlaps. So, the phantom particle is now not exactly a ``phantom'', but is still assumed to follow the motions of its head and tail particles. Thus the prescribed active forces are not affected by the presence of surrounding swimmers, in contrast to the model with the switch-off rule. The new model therefore provides an opposite limiting approach to handling the flagellar-body interaction, thus complementing the study in the main text.
The new model also moves towards the physics of squirmers, whose prescribed surface velocity is likewise not affected by the surrounding squirmers. 

\begin{figure}[h] 
\includegraphics[width=0.375\textwidth]{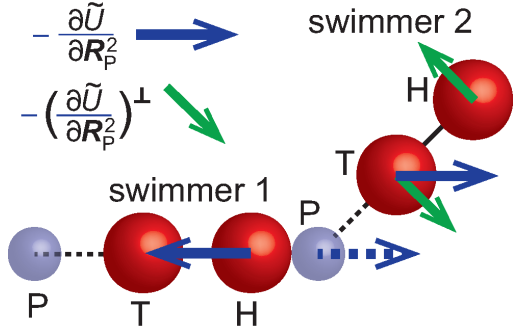} 
\caption{(Color online) 
Schematic of the repulsive forces due to the additional interactions. }
\label{Fig12}
\end{figure}

The potential energy considered here is given by    
\begin{eqnarray}
{\tilde U}\{{\mbox{\boldmath$R$}}_{m}^{\alpha}\} &=& \sum_{\alpha} w(|{\mbox{\boldmath$R$}}^{\alpha}_{\rm H}-{\mbox{\boldmath$R$}}^{\alpha}_{\rm T}|) \nonumber \\ 
&&+\frac{1}{2}\sum_{\alpha\ne \alpha'} \sum_{m,m'} u_{mm'}(|{\mbox{\boldmath$R$}}^{\alpha}_{m}-{\mbox{\boldmath$R$}}^{\alpha'}_{m'}|) \nonumber \\
&&~~~~~~~~~~~~ (m,m'= {\rm H,T,P}), 
\label{total_energy2}
\end{eqnarray}
where $w(r)$ is the intraswimmer potential given by Eq. (\ref{interactionSD}) and the interswimmer repulsive potential $u_{mm'}(r)$ is set to have the same functional form as that of $u(r)$ [Eq. (\ref{interaction})] as
\begin{eqnarray}
u_{mm'}(r) = E_2\biggl(\dfrac{a_{m}+a_{m'}}{r}\biggr)^{24},  
\end{eqnarray} 
where $a_{\rm H}=a_{\rm T}=a$ and $a_{\rm P}=b$. Note that ${\tilde U}$ is given as the sum of $U$ [Eq. (\ref{total_energy})] and the repulsive interactions involving the phantom particles. Since ${\mbox{\boldmath$R$}}^{\alpha}_{\rm P}$ is not an independent variable but is uniquely determined by ${\mbox{\boldmath$R$}}^{\alpha}_{\rm H}$ and ${\mbox{\boldmath$R$}}^{\alpha}_{\rm T}$ according to Eq. (\ref{phantom}), the force is given by 
\begin{eqnarray}
{{\mbox{\boldmath$F$}}_{\alpha,{m}}} &=& - \dfrac{\partial {\tilde U}}{\partial {\mbox{\boldmath$R$}}^{\alpha}_{m}} \nonumber \\ 
&=& - \dfrac{\partial {U}}{\partial {\mbox{\boldmath$R$}}^{\alpha}_{m}}-\dfrac{\partial {\mbox{\boldmath$R$}}^{\alpha}_{\rm P}}{\partial {\mbox{\boldmath$R$}}^{\alpha}_{m}} \cdot \dfrac{\partial {\tilde U}}{\partial {\mbox{\boldmath$R$}}^{\alpha}_{\rm P}}~~(m={\rm H,T}), 
\end{eqnarray}
where the second term of the last line is due to the additional repulsive interactions, whose explicit forms are 
\begin{eqnarray}
\Delta {{\mbox{\boldmath$F$}}_{\alpha,{\rm H}}} &\equiv& - \dfrac{\partial {\mbox{\boldmath$R$}}^{\alpha}_{\rm P}}{\partial {\mbox{\boldmath$R$}}^{\alpha}_{\rm H}} \cdot \dfrac{\partial {\tilde U}}{\partial {\mbox{\boldmath$R$}}^{\alpha}_{\rm P}} \nonumber \\
   &=& \dfrac{\ell_1}{|{\mbox{\boldmath$R$}}^{\alpha}_{\rm H}-{\mbox{\boldmath$R$}}^{\alpha}_{\rm T}|}\biggl(\dfrac{\partial {\tilde U}}{\partial {\mbox{\boldmath$R$}}^{\alpha}_{\rm P}}\biggr)^\bot,  \label{add_head} \\
\Delta {{\mbox{\boldmath$F$}}_{\alpha,{\rm T}}} &\equiv& - \dfrac{\partial {\mbox{\boldmath$R$}}^{\alpha}_{\rm P}}{\partial {\mbox{\boldmath$R$}}^{\alpha}_{\rm T}}  \cdot\dfrac{\partial {\tilde U}}{\partial {\mbox{\boldmath$R$}}^{\alpha}_{\rm P}} \nonumber \\
   &=& -\dfrac{\partial {\tilde U}}{\partial {\mbox{\boldmath$R$}}^{\alpha}_{\rm P}} -\dfrac{\ell_1}{|{\mbox{\boldmath$R$}}^{\alpha}_{\rm H}-{\mbox{\boldmath$R$}}^{\alpha}_{\rm T}|}\biggl(\dfrac{\partial {\tilde U}}{\partial {\mbox{\boldmath$R$}}^{\alpha}_{\rm P}}\biggr)^\bot.  \label{add_tail} 
\end{eqnarray}
Here, 
\begin{eqnarray}
\biggl(\dfrac{\partial {\tilde U}}{\partial {\mbox{\boldmath$R$}}^{\alpha}_{\rm P}}\biggr)^\bot= \bigl({\stackrel{\leftrightarrow}{\mbox{\boldmath$\delta$}}} - \hat{\mbox{\boldmath$n$}}_{\alpha} \hat{\mbox{\boldmath$n$}}_{\alpha} \bigr)\cdot \dfrac{\partial {\tilde U}}{\partial {\mbox{\boldmath$R$}}^{\alpha}_{\rm P}}
\end{eqnarray}
is the component of ${\partial {\tilde U}}/{\partial {\mbox{\boldmath$R$}}^{\alpha}_{\rm P}}$ perpendicular to the swimmer's axis, where ${\stackrel{\leftrightarrow}{\mbox{\boldmath$\delta$}}}$ is the unit tensor and $ \hat{\mbox{\boldmath$n$}}_{\alpha}$ is the unit vector along the swimming axis of the $\alpha$ th swimmer. The physical meaning of these additional forces $\Delta {{\mbox{\boldmath$F$}}_{\alpha,m}}$ ($m=$H,T) is as follows. The tail and phantom particles interact via hypothetical stretching and bending potentials, which transmit a repulsive force acting on the phantom particle; in the tight-binding limit of these interactions, this force is immediately transformed into the translational and rotational forces acting on the head and tail particles (body part) according to Eqs. (\ref{add_head}) and (\ref{add_tail}). A schematic of this situation is shown in Fig. \ref{Fig12}. Note that the present treatment also conserves the total momentum. 

In this Appendix, similarly to the analysis performed in Sec. IV, we investigate the volume fraction dependence of the mean motility and its influence on the clustering behavior with the above-introduced additional interactions (and without the dynamical rule). Here, we use the same parameters and box size as those in Sec. IV. 

\begin{figure}[bht] 
\includegraphics[width=0.48\textwidth]{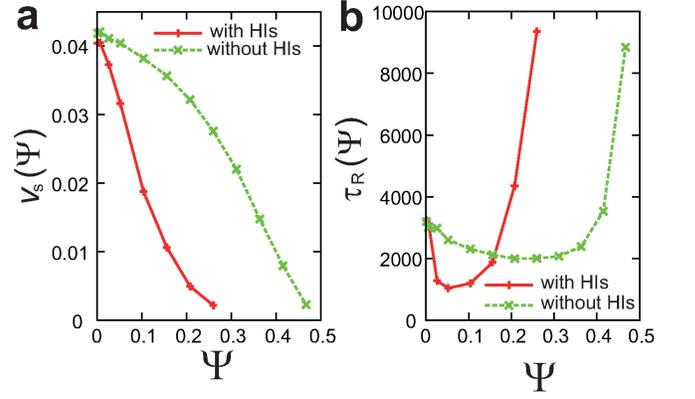} 
\caption{(Color online) 
The average swim speed (a) and the rotational relaxation time (b) for various volume fraction $\Psi$ at $f_{\rm act}=14$. The red and green curves are for the cases with and without HIs, respectively. These curves exhibit similar behaviors to those with the dynamical rule shown in Fig.~\ref{Fig10} but with an overall shift to higher $\Psi$ (due to the absence of the rule-induced slowing down). 
}
\label{Fig13}
\end{figure} 
 
In Figs.~\ref{Fig13}(a) and \ref{Fig13}(b), we plot the dependence on the volume fraction $\Psi$ of the average swimspeed, $v_{\rm s}(\Psi)$, and the rotational relaxation time, $\tau_{\rm R}(\Psi)$, respectively. Similarly to the result shown in Sec. IV [Fig.~\ref{Fig10}(a)], with HIs, $v_{\rm s}(\Psi)$ exhibits a much faster decay than that without HIs. The rotational relaxation also exhibits a similar behavior to that shown in Fig.\ref{Fig10}(b). These observed differences in the (mean) motility for systems with and without HIs can be again attributed to the long-range and cooperative natures of HIs (see the discussion in Sec. IV). Notice that the reduction of $\tau_{\rm R}(\Psi)$ at lower $\Psi$ is more significant than that observed in Fig. \ref{Fig10}(b). This may be because of both the addition of the repulsions and the removal of the dynamical rule, which enhance the collision-induced transfer of the angular momentum. 

\begin{figure}[hbt] 
\includegraphics[width=0.475\textwidth]{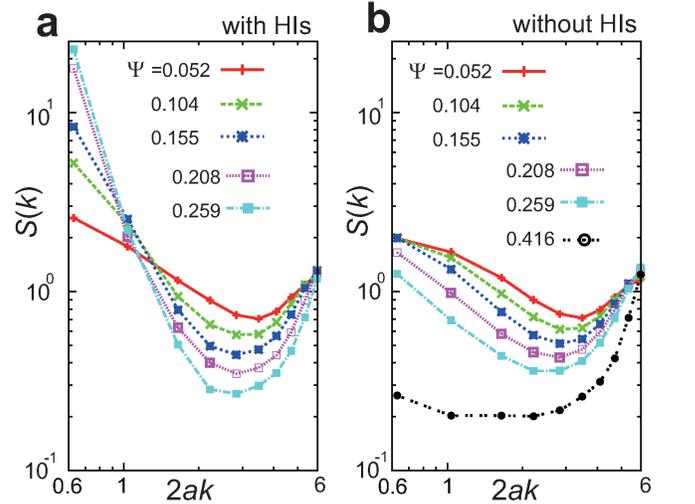} 
\caption{(Color online) 
The structure factor with (a) and without (b) HIs for various $\Psi$ at $f_{\rm act}=14$. With HIs, within the range of $\Psi$ investigated here, the structure factor at small $k(= L/2\pi)$ increases with increasing $\Psi$.}
\label{Fig14}
\end{figure} 

As is discussed in Sec. IV, these differences in the volume fraction dependence of the motility with and without HIs should affect the clustering behavior. In Fig. \ref{Fig14}, we show the structure factor for various $\Psi$ with and without HIs. Without HIs, the weak clustering occurs at smaller volume fraction ($\Psi\lesssim 0.2$) but is highly suppressed at higher $\Psi$. 
From the data of $v_{\rm s}(\Psi)$, the condition $dv_{\rm s}/d\Psi<- v_{\rm s}/\Psi$ holds for $\Psi\gtrsim 0.3$, so that according to the theory \cite{Tailleur_Cates,Cates_Tailleur,CatesR}, the effective compressibility is expected to be negative, leading to the instability of the homogeneous state. However, as noticed in Sec. IV, for the present dumbbell system, at such relatively large volume fractions, direct repulsions (which are omitted in the theory) give a dominant contribution to the compressibility (or pressure). This stabilizes the homogeneous state.  Although the structure factor is found to grow significantly on increasing the magnitude of the active force, further numerical investigations are necessary to clarify the detailed conditions for bulk phase separation, which remain the subject of a future study. 
On the other hand, with HIs, as seen in Fig.~\ref{Fig14}(a), the clustering becomes enhanced with increasing $\Psi$. The fitting of $v_{\rm s}(\Psi)$ represented in Fig.~\ref{Fig13}(a) suggests that the spinodal condition $dv_{\rm s}/d\Psi<- v_{\rm s}/\Psi$ of Refs. \cite{Tailleur_Cates,Cates_Tailleur,CatesR} is satisfied for $\Psi\gtrsim 0.1$. Notice that the clustering tendency at relatively larger volume fractions is much more enhanced than that exhibited by the model with the switch-off rule [for comparison, see Fig. \ref{Fig11}(a)]. In the model used here, the active forces of the swimmers are not reduced at close proximity, and therefore the activity-induced hydrodynamic attractions are strong enough to overcome repulsive interactions even for larger $\Psi$.

In this Appendix, by performing an analysis similar to that in Sec. IV, we have explored the clustering behavior of the dumbbell model with a different treatment of the phantom particle. Without HIs, some marked differences arise between the two treatments; that is, the collective behaviors are strongly dependent on the local rules or interactions. On the other hand, because the long-range and non-local natures of HIs govern the collective dynamics, the global picture regarding the hydrodynamic effects is not much altered. This robustness to the local interaction rules is an important result of the present study. 
The above simulation results, along with those shown in Sec. IV, show that theory can qualitatively explain the characteristic features of the effective attraction caused by the density dependent motility, although there are some discrepancies between the theory \cite{Tailleur_Cates,Cates_Tailleur,CatesR} and the present simulation, especially at higher volume fractions, where some effects omitted from the theory (for example, steric repulsions due to the excluded volume) can come to dominate. Although the effects of long-range HIs are largely separable from near-field and collisional effects, the latter are separately important, at least at the quantitative level. Here one expects strong dependencies both on particle shape and on whether the propulsion is described by a force-prescribed or a velocity-prescribed mechanism.


\begin{thebibliography}{170}
\bibitem{comment2}
In the original works in Refs. \cite{Graham1,Graham2,Gyrya_Aranson_Berlyand_Karpeev}, a microorganism is represented as a set of point forces. However, in the simulation of the continuum model using a discrete lattice, a force should be given through a prescribed spatial distribution with a finite width, which must be larger than the lattice spacing.
\bibitem{comment3}
The factor, $\int d{\mbox{\boldmath$r$}} \phi^2_{\rm T}/\Omega_{\rm T}=0.59$, arises from the definition of the particle velocity [Eq. (\ref{particle_velocity})]. In the Brownian simulations, the active force is multiplied by this factor to match the swim-speed of an isolated swimmer with and without HIs.
\bibitem{comment1}
In the present FPD method, ${\mathcal F}^{\rm vis}_{m}= \int d{\mbox{\boldmath$r$}} (\phi_{m}/\Omega_{m})[\nabla\cdot\eta(\nabla{\mbox{\boldmath$v$}}^\dagger+\nabla{\mbox{\boldmath$v$}}) ]^\bot$, ${\mathcal F}^{\rm sp}_{m}= -\int d{\mbox{\boldmath$r$}} (\phi_{m}/\Omega_{m}) (\phi_m {\partial w(| {\mbox{\boldmath$R$}}_{\rm H}- {\mbox{\boldmath$R$}}_{\rm T}|)}/{\partial {\mbox{\boldmath$R$}}_{m}})^\bot$, and  ${\mathcal F}^{\rm act}_{\rm T}=  \int d{\mbox{\boldmath$r$}} (\phi_{\rm T}/\Omega_{\rm T}) (\phi_{\rm T} f_{\rm act} \hat{\mbox{\boldmath$n$}})^\bot$ 
are the viscous, the constraining, and the active forces acting on the particle, respectively.
\bibitem{comment4} 
Similar hydrodynamic instabilities were also theoretically argued by Baskaran and Marchetti \cite{Baskaran_Marchetti}. However, in their study, the growth of density fluctuations is predicted to occur for pullers, but it does not for pushers, which is different from the present results. This diffrence may be due to the lack of the near-field hydrodynamic effects in the analysis of Ref. \cite{Baskaran_Marchetti}.
\end{thebibliography}
\end{document}